\documentclass[a4paper,aps,prb,reprint,showpacs]{revtex4-1}
\usepackage{CJK}
\usepackage{amsmath,amssymb,mathrsfs,amsbsy}
\allowdisplaybreaks
\usepackage[pdftex]{graphicx}
\usepackage{float} 
\usepackage{xcolor}
\usepackage{booktabs} 
\usepackage{hyperref}

\DeclareMathOperator{\csch}{csch}

\begin{document}

\bibliographystyle{unsrt}

\title{Exact correlation functions at finite temperatures in Tomonaga-Luttinger liquid with an open end}

\author{Naira Grigoryan}
\affiliation{Institute of Fundamental Technological Research, Polish Academy of Sciences, Adolfa Pawińskiego 5b, 02-106 Warsaw, Poland}

\author{Piotr Chudzinski}
\affiliation{Institute of Fundamental Technological Research, Polish Academy of Sciences, Adolfa Pawińskiego 5b, 02-106 Warsaw, Poland\\
and\\
School of Mathematics and Physics, Queen's University Belfast, University Road, Belfast, NI BT7 1NN, United Kingdom}

\date{\today}

\begin{abstract}

The paradigmatic state of a 1D collective metal, the Tomonaga-Luttinger liquid (TLL), offers us an exact analytic solution for a strongly interacting quantum system not only for infinite systems at zero temperature but also at finite temperature and with a boundary. Potentially, these results are of high relevance for technology as they could lay the foundation for a many-body description of various nanostructures. For this to happen, we need expressions for local (i.e., spatially resolved) correlations as a function of frequency. In this study, we find such expressions and study their outcome. Based on our analytic expressions we are able to identify two distinct cases of TLL which we call Coulomb metal and Hund metal, respectively. We argue that these two cases span all the situations possible in nanotubes made out of p-block elements. From an applications viewpoint, it is crucial to capture the fact that the end of the 1D system can be coupled with the external environment and emit electrons into it. We discuss such coupling on two levels for both Coulomb and Hund metals: i) in the zeroth order approximation, the coupling modifies the 1D system's boundary conditions; ii) stronger coupling, when the environment can self-consistently modify the 1D system, we introduce spatially dependent TLL parameters. In case ii) we were able to capture the presence of plasmon-polariton particles, thus building a link between TLL and the field of nano-optics. 

\end{abstract}

\maketitle

\section{Introduction}


One-dimensional (1D) metals host physics that is quite different from what we are used to in standard bulk 3D materials that surround us every day. It was recognized already back in the 1950s by Tomonaga \cite{tomonaga1950remarks} and Luttinger \cite{luttinger1963exactly} that 1D electrons have purely collective properties, then the model was solved by Mattis and Lieb \cite{10.1063/1.1704281}, who provided an exact solution to Luttinger's formulation. Later, this work was gradually extended by computing detailed correlation functions and spectral properties, including finite-temperature effects. \cite{cross1979new, schulz1983quantum, PhysRevB.47.6740, PhysRevB.46.15753, SCHONHAMMER1993225, Bourbonnais2001}. In 1981, Duncan Haldane established Tomonaga Luttinger liquid (TLL) as a new paradigm \cite{haldane1981luttinger} - an alternative to Landau-Fermi liquid that describes 1D metals. Overall the behaviour of the bulk 1D metal is now quite well understood: we know that there is going to be an interaction-induced pseudo-gap in the spectrum, sometimes called zero-bias-anomaly \cite{egger2001bulk}. The beautiful thing is that this non-trivial model, dominated by many-body effects, admits an exact analytic solution. This works also at finite temperatures and, as shown in seminal works of Affleck and Ludwig \cite{affleck1997boundary, affleck1991kondo, affleck1992exact, affleck1993exact}, close to the edges of a low dimensional system. In the boundary problem, at zero temperature, the spectral function is going to be a power law with an exponent changing as the distance away from the boundary varies. These boundary-induced features are particularly important for realistic applications in nanostructures where the surface-to-volume ratio is large. 

Over the past two decades, we have witnessed an intense development in the field of low-dimensional nanostructures. This progress has been mostly experimental/technological, and it led to the need of developing their theoretical description. The TLL formalism is particularly useful, as it enables to incorporate influence of many-body interactions also in finite temperatures. To bridge the gap with real-life devices, such as field electron emitters, it is essential to capture how such many-body systems will behave when they are exposed to the outside world. 
While TLL provides a complete description of an infinite-size 1D system it remains an active field of research to answer how these quantum field theory analytic solutions will behave when coupled to the environment. Therefore, we aim to derive an exact analytic formula for the correlation function in the technologically relevant case when the boundary of a 1D system interacts with its environment, in particular when either there is an intense fermionic emission from the end or when the TLL interacts with a local plasmon-polariton state at its end.

One remarkable outcome of Affleck and Ludwig's work, based on boundary conformal field theory, is the proof that the spectrum on the edge of a 1D system can contain discrete boundary states.
These findings are supported by exact solutions from the nonlinear-sigma model
\cite{ng1994edge} and many numeric calculations, mostly by means of DMRG 
\cite{white1993numerical, qin1995edge}. In particular, it was shown that boundary modes can exist in critical metals as well
\cite{nataf2021edge, fath2006logarithmic}.

While the theory can provide a qualitative picture, and numerics can substantiate it quantitatively, one wonders if any exact analytical solutions are available for the experimentally most pertinent situation: a boundary and a finite temperature in the energy/frequency domain. Unfortunately, the exact \emph{analytic} solution is known only in the real (space-time) domain, while most experiments, for instance, STM can measure signal in frequency (energy) space. Thus, we need to perform a (partial) Fourier transform: we need to make a Fourier transform in the time domain while keeping the information about the distance from the edge. It is worth noting that in our problem, with translational invariance broken, we have a relative real-space coordinate and an absolute distance from the edge $r$. The first variable is taken to be constant and small (at least in the case of e.g. an STM probe) while the second one is the variable of our interest here.

There are not too many examples of such Fourier transforms. This is unfortunate because the TLL's real-space correlation functions contain singularities which make their numerical integration tricky. Facing this challenge we decided to obtain such Fourier transform not for the case of arbitrary value of TLL's compressibilities (i.e. arbitrary strength of electron-electron interaction), but instead to identify \emph{specific} values of compressibilities when the transform is possible. In our studies, we were able to identify two classes of 1D systems with distinct local densities of states (LDOS) in the vicinity of the boundary: one which we call Coulomb metal, and the other - Hund metal, and then compare the two. As we shall see, these two cases represent the two main types of boundary, i.e., with and without the boundary resonance; thus they suffice as two examples capturing the physics of the boundary problem. We emphasize here that a very substantial number of past works have already allowed researchers to understand the physics of the boundary problem quite well. So our aim in this work is solely to provide examples of exact analytical solutions together and extend them to the situation when the coupling with an environment is so strong that plasmon-polariton physics appears. 

The outline of this work is as follows: in Sec.II, we introduce the model, define Coulomb metals and Hund metals, and provide a few examples of their realizations. In Sec.III, we explain in detail how the correlation functions are computed in real space and emphasize changing the boundary condition changes this result. The last issue is frequently overlooked and set in a standard way, which as will be seen, is only one of several options. Then in Sec.IV, we perform the aforementioned LDOS's Fourier transform (for two cases) and illustrate the outcome. In Sec.V, we perform the same steps to obtain the dominant contribution to TLL's charge susceptibility, also in the analytic form. In Sec.VI, we show how to modify our formalism to capture the formation of the plasmon-polariton state, which is the strong coupling limit of radiation. Our results and their implications are summarized in Sec.VII.

\section{Model}

The full Hamiltonian of the system consists of bulk and boundary terms:
\begin{equation}
    H_{tot}=H_{TLL}+H_{boun}
\end{equation}
we will first define the 1D bulk term $H_{TLL}$ as this determines what fields have to be used to define the boundary term $H_{boun}$.

Tomonaga-Luttinger liquid is a prototype of a strongly interacting metal. It offers us an opportunity to explore in a non-perturbative way how (Hartree, i.e. density-density) interactions modify the properties of the system, in particular how they modify the distribution of electrons. The Hamiltonian of this collective state reads:

\begin{equation}\label{eq:ham-TLL}
       H_{TLL} =  \sum_{\nu}^{\bar{N}}\int dx \left[\frac{u_{\nu}}{K_{\nu}^2} (\nabla \phi_{\nu}(x))^2 + u_{\nu}K_{\nu}^2(\nabla \theta_{\nu}(x))^2\right]
\end{equation}

where $\bar{N}$ is a number of bosonic modes in our system. We have imposed the same convention for $K_{\nu}$ parameters like in seminal works on TLL boundaries by Eggert and Affleck, the Ref. [\onlinecite{eggert1992magnetic, PhysRevB.56.15615}]. It has to be emphasized that this is different than the usually used convention, the two are related as $K_{usual}^{1/2}\rightarrow K_{Eggert}$. The functional form of all correlation functions stays the same, the non-interacting point $K=1$ remains and the only change is the power with which $K_{\nu}$ parameters enter the expressions. We decided to use the convention of Ref. [\onlinecite{PhysRevB.56.15615}] because we frequently compare our results with that original work where boundary correlation functions were derived for TLL. 

The form of Hamiltonian given in Eq.\ref{eq:ham-TLL} immediately informs us that the physics is written in terms of collective, plasmon-like modes related to density $\phi_{\nu}$ and momentum density $\theta_{\nu}$ of the $\nu$-th mode. These modes are also related to the sum and difference of excited TLL bosons $b^{\dag}_{\pm q,\nu}$, respectively. The physics of the system is encoded in the values of velocities $u_{\nu}$ and TLL parameters $K_{\nu}$, the latter ones are proportional to the compressibilities of respective bosonic modes. When Galilean invariance is obeyed the following relation holds $u_{\nu}K_{\nu}=V_F$, thus the two quantities are connected, varying one implies immediately a variation of another. In the non-interacting case, we have $K_{\nu}=1$, while the presence of interactions shifts this value either downwards (for repulsive interactions) or upwards (for attractive interactions). The number of modes depends on the number of fermionic constituents of the model: for a spin-full 1D chain, there are two modes: $\nu=\rho,\sigma$, while for a two-leg-ladder (which is also an appropriate model for nanotubes), there are four modes: $\nu=\rho\pm,\sigma\pm$ (where $\pm$ represent symmetric and anti-symmetric fluctuations in the two legs of the ladder), and so on towards more complicated systems with a larger number of constituents $\bar{N}$. In the following, we shall try to keep our reasoning as general as it can be, resorting to specific cases only for illustrative purposes.

On the top of this bulk Hamiltonian, describing collective modes of a 1D strongly correlated system, we add a boundary term:
\begin{equation}\label{eq:ham-bound}
    H_{boun}=V_0 \prod_{\nu}^{\bar{N}} \cos\vartheta_\nu(x=0)
\end{equation}
where $V_0$ is a large amplitude, that imposes the boundary. This is a local term: a single-particle term at the boundary, at $x=0$. Being a single particle operator, it necessarily contains bosonic fields of all modes. The choice of its fields' content depends on the physical nature of the boundary. For instance, the list of all possible (sixteen) single-particle backscattering terms for the two-leg ladder has been given in Ref. [\onlinecite{wu2003competing}]. Physically, if the boundary imposes a given value of density (e.g. spin density), then the boundary condition is imposed on the $\phi_{\nu}(x=0)$ field, that is $\vartheta_{\nu=\sigma}=\phi_{\sigma}$ in Eq.\ref{eq:ham-bound}, but if spin-flips are promoted, then the boundary condition is imposed on $\theta_{\nu}(x=0)$, which is in our example  $\vartheta_{\nu=\sigma}=\theta_{\sigma}$ in Eq.\ref{eq:ham-bound}. Mathematically, due to the particle number conservation law, the first case is defined as a standard Dirichlet boundary condition (BC) set for $\phi_{\nu}(x=0)$, contrary to which the second one is defined in terms of momentum fields $\theta_{\nu}(x=0)$  Neumann boundary condition (sometimes called radiative boundary conditions [\cite{PhysRevB.58.10761}], especially if entire fermions are being removed by the BC operator). We will elaborate on this modification in the following section, Sec.\ref{ssec:bound-cond-def}, but here we note that the first case is applied more frequently, even in textbooks, thus sometimes being considered as the only option. For the illustrative purposes in the paper, we shall then use the second choice, looking at the odd character of Neumann BC.

Coulomb metal will be the case when only the charge mode compressibility will be different than one, $K_{\rho+}\neq 1$, and it can be different than one by a large margin. The values for all other parameters will stay close to one. This can happen in the case of a material dominated by long-range Coulomb interactions, hence the name. One example is a free (un-bundled) carbon nanotube where interactions are only weakly screened. These are usually modeled by two two-leg ladder Hamiltonian (or multi-leg ladder for MWCNT), but there is always a single, totally symmetric charge mode $\rho+$ and it is the one that will be affected by interactions.

Hund metal will be the case when both charge and spin modes are subjected to long-range interactions. This may be the case in a multi-orbital system which is in an orbitally selective Mott phase. The Mott insulating bands will provide a magnetically ordered background (usually anti-ferromagnetic) which upon averaging out, will provide interaction-mediating bosons for the remaining metallic state. The orbital that remains metallic is usually split by the Hund coupling, hence the name we have chosen. Since the same bosons are mediating charge and spin interactions, $K_{\rho}$ and $K_{\sigma}$ are expected to be approximately equal. 

In nanostructures, an unavoidable presence of boundaries breaks the translational invariance, and it is definitely worth investigating what electron distribution in their vicinity is. For the non-interacting system, one expects homogeneous distributions from Bloch waves plus Friedel oscillations that always decay like $1/x^1$ in 1D. Including interactions can make this simple picture much more complicated and interesting. Uncovering this phenomenology in the frequency space, as the coherent excitations are competing with thermal and bath effects, is the objective of this paper. 

\section{Correlation functions}

\subsection{Past results}

\paragraph{Bulk TLL} A great advantage of working with TLL is that all correlation functions can be in principle computed exactly. A quantity of our interest is the spectral function, i.e., an imaginary part of the retarded Green's function which in TLL is obtained by an analytic continuation of a time-ordered average anti-commutator $\langle[]_{+}\rangle$ of fields. For instance, the result for the zero temperature spectral function in an infinite TLL is well-known [\cite{giamarchi2003quantum}] and reads:

\begin{widetext}
\begin{multline}\label{eq:bulkTLL0}
    A_0(x,t; T=0) = -i Y(t) \langle[\psi_{rs}(x,t), \psi_{rs}^{\dag} (0,0)]_{+}\rangle = - i \frac{Y(t)}{2 \pi} e^{i r k_{F} x} \lim_{\epsilon \rightarrow 0} \Biggl\{ \frac{\alpha_{cut-off}+i(v_F t - rx)}{\epsilon+i(v_F t - rx)}\\
    \times \prod_{\nu = \rho, \sigma} \frac{1}{\sqrt{\alpha_{cut-off}+ i (u_\nu t -rx)}} \Biggl( \frac{\alpha_{cut-off}^2}{(\alpha_{cut-off}+i u_\nu t)^2+x^2} \Biggl)^{\bar{\gamma_{\nu}}} +  \begin{pmatrix} x \rightarrow -x \\ t \rightarrow -t \end{pmatrix}  \Biggl\}
\end{multline}
\end{widetext}
where $Y(t)$ is the step function (that imposed causality of the original time-ordered function), $\alpha_{cut-off}$ is the UV cut-off of the theory and the exponent $\bar{\gamma_{\nu}}$ that comes from $\phi_{\nu}$, and $\theta_{\nu}$ fields is: 

\begin{equation}
    \bar{\gamma_{\nu}} = \left( K_{\nu} + K_{\nu}^{-1} -2 \right) /8 >0
\end{equation}

Our study is in essence about finding Fourier transforms of the $A(x,t)$ expressions. In this $T=0$ bulk case, which is simply a power law, one finds straightforwardly the following Fourier transform:

\begin{equation}
    A_{R,s} (q, \omega)  \sim (\omega - u_{\sigma} q)^{\zeta-1/2} |\omega - u_{\rho} q|^{(\zeta-1)/2} (\omega+v_{\rho} q)^{\zeta/2}
\end{equation}

where $\zeta=\sum_{\nu}\bar{\gamma_{\nu}}$, so we see that interactions enter, through $K_{\nu}$, in a highly non-trivial, non-perturbative way -- they modify exponents of the power law. This implies that the result is non-perturbative. The further advantage of TLL is that it obeys conformal invariance (CFT), hence the correlation functions remain accessible at finite temperature (and finite size) as well. The finite temperature expressions can be found through CFT mapping of a plane on a cylinder (thus imposing a periodic condition in time) which is equivalent to a substitution $r\rightarrow sinh(r/\beta)$ (here $\beta$ is an inverse temperature). In comparison with Eq. \ref{eq:bulkTLL0}, the formulas are more complicated since they contain hyperbolic functions, but analytical expressions in real space are still possible to write. 

\paragraph{TLL with a boundary} The real space expressions for correlation functions have been also obtained, but only in an integral form, for a finite temperature TLL with a boundary in Ref. [\onlinecite{PhysRevB.56.15615}]. They defined the following total time ordered Green's function (the last term gives Friedel oscillations):
\begin{widetext}

\begin{equation}
N(x,y,t;\beta)=Im[\langle[\Psi_{rs}(x,t), \Psi_{rs}^{\dag} (y,0)]_{+}\rangle]=Im[G_{RR}(x,y,t;\beta)+G_{LL}(x,y,t;\beta)+\cos(i k_F(x+y))G_{LR}(x,-y,t;\beta)]
\end{equation}
where
\begin{equation}
\Psi(x,t)=\exp(ik_Fx)\psi_R(x,t)+\exp(-ik_Fx)\psi_L(x,t) ~ and ~ G_{RR}(x,y,t;\beta)=\langle \psi^{\dag}_R(x,t) \psi_R(y,0)\rangle
\end{equation}

The result of our reasoning will be a function $N(\omega, r; \beta)$ of frequency (energy) and distance from the boundary with inverse temperature $\beta$ as a parameter. The  $N(x=y=r,t;\beta)$'s Fourier transform is an integral transform $\mathfrak{F}[t\rightarrow\omega]$ of a real-time expression which consists out of two parts, one obtained through \textbf{BP} described in the box below, and the zero temperature regularization. It reads as follows:

\begin{multline}\label{eq:boundTLL}
    N(\omega, \beta, r) = \frac{2}{\alpha_{cut-off} \pi^2} v_c^{-a_c} v_s^{-a_s} \int_{0}^{\infty}dt\cos{\gamma (t)}
    \Bigg[ \cos{\omega t} \left( \frac{\sinh{\frac{\pi}{\beta} t}}{\frac{\pi}{\beta} \alpha_{cut-off}}\right)^{-a_s-a_c} \\ \times \Bigg|  \frac{\sinh{\frac{\pi}{v_c \beta}} (2r+v_c t) \sinh{\frac{\pi}{v_c \beta}} (2r-v_c t)}{\sinh^2  \frac{2 \pi r}{v_c \beta}}\Bigg|^{-b_c/2} 
    \Bigg| \frac{\sinh{\frac{\pi}{v_s \beta}} (2r+v_s t) \sinh{\frac{\pi}{v_s \beta}} (2r-v_s t)}{\sinh^2  \frac{2 \pi r}{v_s \beta}}\Bigg|^{-b_s/2}   \\ -\left(\frac{t}{\alpha_{cut-off}}\right)^{-a_s-a_c} \Bigg| 1-\left(\frac{v_c t}{2 r}\right)^2 \Bigg|^{-b_c/2} \Bigg| 1-\left(\frac{v_s t}{2 r}\right)^2\Bigg|^{-b_s/2}   \Bigg]
\end{multline}


where the $\gamma(t)$ is a phase shift function (written explicitly in App.A), and the exponents $a_i$ and $b_i$ depend on $K_{\nu}$, as we will show in the next subsection. In general, this formula describes the case of any TLL that has \emph{two} linear dispersions: one can have any number $\bar{N}$-modes and the formula Eq.\ref{eq:boundTLL} will remain valid provided they split in exactly two sets -- there are $m$ (degenerated) modes with velocity $v_s$ and $\bar{N}-m$ modes with velocity $v_c$. In real space, it is possible to find a formula for a greater number of bosonic dispersions, but finding a Fourier transform of those is beyond the scope of our work here. The above expression, Eq.\ref{eq:boundTLL}, has been obtained using a method to evaluate TLL correlations with a boundary originally derived by Affleck and Eggert [\cite{eggert1992magnetic}] and described in the box below. 

\fbox{
\begin{minipage}{0.95\textwidth}
The boundary procedure, \textbf{BP}, involves several transformations of the bosonic fields, namely the fermionic field undergoes the following series of mappings:

Initially the relation between fermionic field $\psi(x)$ and bosonic fields $\phi,\theta$ reads as usual:

   $$ \psi_{R} \left( x_{1}, \tau_{1} \right)  \varpropto e^{-i \left[ \phi  \left( x_{1}, \tau_{1} \right) -\theta  \left( x_{1}, \tau_{1} \right)  \right]  } $$

\begin{description}
    \item[Step 1] re-scale $\phi$ fields by $K$ and $\theta$ fields by $1/K$ (this is to move to non-interacting fields $\tilde{\phi}$, to avoid the non-local interactions later on)

    $$ \psi_{R} \left( x_{1}, \tau_{1} \right)  \varpropto e^{-i \left[K \tilde{\phi}  \left( x_{1}, \tau_{1} \right) -\frac{1}{K} \tilde{\theta}  \left( x_{1}, \tau_{1} \right)  \right]  } $$
    
    \item[Step 2] move to basis of chiral $\tilde{\phi}_{L,R}$ fields

    $$ \psi_{R} \left( x_{1}, \tau_{1} \right)  \varpropto e^{-i \left[K \left( \frac{\phi_{L} - \phi_{R}}{2}\right) -\frac{1}{K} \left( \frac{\phi_{L} + \phi_{R}}{2}\right)  \right]  \left( x_{1}, \tau_{1} \right) } $$

      \item[Step 3] when the boundary condition is set by large $\cos\phi$, thus by pinning $\phi(x=0)=0$, in terms of chiral fields this translates to a non-local condition
    
    $$\phi_{L}(x)=\phi_{R}(-x)$$ 
upon which the fermion field is transformed as:

    $$ \psi_{R} \left( x_{1}, \tau_{1} \right)  \varpropto e^{-\frac{i}{2} \left[ \left( K -\frac{1}{K} \right) \phi_{R}^{e} \left( - x_{1}, \tau_{1}  \right) -  \left( K -\frac{1}{K} \right) \phi_{R}^{e} \left( x_{1}, \tau_{1}  \right)\right]} $$
    
    \item[Step 4] one can now move back from chiral fields to $\tilde{\phi},\tilde{\theta}$ fields

    $$ \psi_{R} \left( x_{1}, \tau_{1} \right)  \varpropto e^{-\frac{i}{2} \left[ \left( K -\frac{1}{K} \right) \left[ \theta^{e} -\phi^{e} \right] \left( - x_{1}, \tau_{1}  \right) -  \left( K -\frac{1}{K} \right) \left[ \theta^{e} -\phi^{e} \right] \left( x_{1}, \tau_{1}  \right)\right]} $$
    
\end{description}
\end{minipage}
}
\end{widetext}

And one is left with computing correlation functions of non-interacting fields $\langle \phi^e(x_1,\tau_1)\phi^e(x_2,\tau_2)\rangle$ which are known. However, in the last line, we see fields at both $x_1$ and $-x_1$. As a result of this, even for equal space correlation functions $x_1=x_2=r$, one obtains correlators $\langle\phi^e(r,t)\phi^e(r,t')\rangle$ as well as $\langle\phi^e(r,t)\phi^e(-r,t')\rangle$. The latter ones produce terms $\propto (2r\pm v_{\nu}t)$, and this $r$ dependence indicates the translational symmetry breaking. The finite temperature correlations are obtained as usual thanks to conformal symmetry transformation $z\rightarrow \beta/2\pi log(z)$ which leads to hyperbolic $sinh(\pi t/\beta)$ functions present in Eq. \ref{eq:boundTLL}. All the above reasoning was performed for a single mode TLL $\phi$ (spinless fermions), but since the fermion field can be factorized into separate contributions of bosonic modes $\phi_{\nu}$, the generalization to the multi-mode case is straightforward: one makes the same steps separately for each mode.  

\subsection{Boundary conditions (BC)}\label{ssec:bound-cond-def}
 
The correlation function of a given operator, for instance, the Green function in bulk still ought to follow Eq. \ref{eq:bulkTLL0}, as it only depends on $a_i$ exponent. This is because in the terms that contain $b_i$ the numerator and denominator can be reduced for large $r$. The relation between the exponents $b_i$ and the TLL parameters is determined by the nature of the boundary condition (BC) itself. If the boundary is set by backscattering potential, the fields $\phi_\nu$ are locked at $x=0$. This is the standard case that has been applied in Ref. [\onlinecite{PhysRevB.56.15615}]. Then we would have $ a_{\nu}= \frac{{K_{\nu}}^2+K_{\nu}^{-2}}{4}$, $b_{\nu}= \frac{K_{\nu}^{-2}-{K_{\nu}}^2}{4}$, and indeed at the boundary the correlation function decays with $1/K_{\nu}^2$ exponent because the fluctuations of $\phi(x=0)$ are frozen. However, as already mentioned in the preceding section, different situations and different boundary conditions at $x=0$ are also possible. We first discuss different variants of backscattering BCs and then we will move to the radiative BC, and then we complete the argument with a discussion of a realistic nanostructure: carbon nanotube.

\paragraph{Various cases for $H_{boun}$.} In Ref. [\onlinecite{PhysRevB.56.15615}], the simplest case of a single mode TLL was considered, and the impurity term $\cos\phi(x=0)$ was responsible for setting the boundary where the density-wave pinning took place\footnote{please note that in Ref. [\onlinecite{PhysRevB.56.15615}] the canonical notation is swapped $\phi\leftrightarrow\theta$.}. For the case of a multi-mode TLL, setting the boundary, i.e. writing explicitly Eq. \ref{eq:ham-bound}, is a non-trivial step that also contains some physics of the problem. The simplest case, that is commonly taken, is to take in Eq. \ref{eq:ham-bound} the case $\vartheta_{\nu}\equiv\phi_{\nu}$ for \emph{all} bosonic modes. Then:
\begin{equation}
    H_{boun}^{usual}=V_0 \prod_{\nu}^N \cos(\phi_{\nu}(x=0))
\end{equation}
which is the pinning of the simplest CDW. In general the boundary Hamiltonian is proportional to the real part of a chosen single-particle operator, namely:
\begin{equation}
    H_{boun}=V_0 Re[\hat{O}_{bd}^{i}(x=0)]
\end{equation}

A two-leg ladder, with a multitude of possible orderings, is a minimal model that can host more rich boundary physics. There are several possible $\hat{O}_{bd}^{i}$, for instance for the two leg-ladder there are eight\cite{wu2003competing, chudzinski2008orbital} possible two-body terms $\hat{O}_{bd}^{i}(x=0)\propto c_{l}^{\dag}(x=0)c_{l'}(x=0)$ (where $l$ is a spin, but also more generally, orbital/leg index), and the nature of boundary scattering determines which one should be chosen. If there is a spin-flip:
$$\hat{O}_{bd}^{s-flip}(x=0)\propto (c_{\uparrow}^{\dag}(x=0)c_{\downarrow}(x=0)+h.c.)$$
or for the carbon-nanotube two leg-ladder also valley-flip
$$
\hat{O}_{bd}^{v-flip}(x=0)\propto (c_{K}^{\dag}(x=0)c_{K'}(x=0)+h.c.)
$$
involved in the process, then some modes will have different boundary conditions than others. Then the boundary term will read:

\begin{equation}\label{eq:standard-with-sf}
H_{boun}^{s-flip}(x=0)\propto V_0 \cos \phi_{\rho \pm} \cos \theta_{\sigma \pm}    
\end{equation}

and
\begin{equation}\label{eq:standard-with-vf}
H_{boun}^{v-flip}(x=0)\propto V_0 \cos \phi_{\rho+} \cos \theta_{\rho-}  \cos \phi_{\sigma+} \cos \theta_{\sigma-}    
\end{equation}

respectively.

For the spin sector, this unusual boundary condition $\hat{O}_{bd}^{s-flip}$ will have the following interpretation: a local Dzyaloshinskii–Moriya (DM) interaction, that is, locally generated spin-flip processes. One can expect that such a term will naturally arise on the edge of a heavy atom chain with nonequivalent (lower symmetry) lattice sites\cite{Deger2020}. Usually, DM interaction approaches zero because contributions from various sites cancel out due to the high symmetry of the crystal lattice. However, this cancellation will not work on the edge of the 1D system where e.g. only one out of two sites is present. Here, we shall then have a strong spin-flip term proportional to $\cos\theta_{\sigma}$ which in the language of original spins, corresponds to $S^{+}, S^{-}$ processes with a large amplitude. 

\paragraph{Radiative boundary condition} One can consider a situation where there is an intense emission of carriers close to the end of a 1D system. Then so-called \emph{radiative} boundary conditions apply. These were first identified in Ref. [\onlinecite{eggert1992magnetic}] and can be written as:
\begin{equation}\label{eq:radiatBC}
(V_{F}g^{-2}\partial_x + \partial_t)\phi(x_0)=p(x_0)
\end{equation}

where $p(x_0)$ is a probability of emitting/injecting carrier at a point $x_0$ and $g$ is interaction amplitude, in the lowest order $g\propto 1-K$.  When $K_{\rho}<0.5$ (such as in our case), then $g\gg V_F$ (please note that $K_{\rho}=0.5$ corresponds to $U\rightarrow\infty$ in the Hubbard model, our $K_\nu$ will be even smaller). Thus, we should focus on the time derivative term in the BC, the Eq.\ref{eq:radiatBC}. We use the central relation of Hamiltonian field theory:
\begin{equation}\label{eq:time-deriv}
    \partial_t \phi = \frac{\delta H_{tot}}{\delta \theta}= \Big(\partial_{\theta}-\nabla\cdot\frac{\partial}{\partial(\nabla\theta)}\Big)H_{tot}
\end{equation}
to arrive at a boundary term for the canonically conjugated field $\theta(x_0)$. This is because $H_{TLL}[\nabla\theta]$ always contains $\nabla\theta$ thus it gives a finite value of a derivative $\frac{\partial}{\partial(\nabla\theta)}$ (from the second term in Eq.\ref{eq:time-deriv}). To compensate for it one needs to choose $H_{boun}[\theta]\propto \cos\theta$ which will produce a finite $\partial_{\theta}$ from the first term in Eq.\ref{eq:time-deriv} and arrive at finite value $p(x_0)$, the stationary point. This sets the boundary condition $\propto\cos(\theta(x_0))$ thus fixes the canonically conjugated momentum field $\theta(x_0)$. 

As a result, the following modification in the above procedure by Affleck and Eggert is necessary: \emph{in step 3 of \textbf{BP} we now impose the boundary condition for the original $\theta$ field. Thus, now for the chiral field, the boundary condition reads:}

\begin{equation}
\phi_{L}(x)=\pmb{-}\phi_{R}(-x)
\end{equation}

The entire reasoning stays the same, but signs need to be changed in front of some coefficients which ultimately results in a change of sign in Eq. \ref{eq-b-def-thet}.
This implies a different relation between $b_i$ exponents and $K_{\nu}$, namely:

\begin{equation}
    a_{s,c}= \frac{{K_{s,c}}^2+K_{s,c}^{-2}}{4}
\end{equation}

\begin{equation}\label{eq-b-def-thet}
    b_{s,c}= \frac{K_{s,c}^{2} - K_{s,c}^{-2}}{4}
\end{equation}

The change of sign in $b_{s,c}$ has physical meaning: since now the $\theta_{\nu}$ fields are frozen on the boundary, the correlation function on the boundary decay with exponent $\sim K_{\nu}^2$

Now let us assume that the end of our 1D system keeps emitting fermions, the radiative boundary condition as introduced in Eq.\ref{eq:radiatBC}. If the emission on the boundary is featureless (full fermion radiation), we choose $\phi_{\nu}\rightarrow \theta_{\nu}$ for the boundary conditions of all bosonic modes. Then:
\begin{equation}\label{eq:ourBC}
    H_{boun}^{}=\mathcal{T}_0 \prod_{\nu}^N \cos(\theta_{\nu}(x=0))
\end{equation}
where $V_0$ is now equivalent to transmission $\mathcal{T}_0$, in either case, it is considered to be a strong cosine perturbation to the bulk TLL that imposes the \textbf{BP} modification of correlation functions. The Eq.\ref{eq:ourBC} gives explicitly the form of boundary used throughout this paper, i.e., the Eq.\ref{eq:ham-bound} is taken with $\vartheta_{\nu}\equiv\theta_{\nu}$ for all bosonic modes $\nu$ unless stated otherwise. For emitting electrons with a spin-flip, the boundary term involves the coupling of charge and spin modes, while for a valley-flip, it captures the valley mixing effects, leading to distinct boundary conditions.

For emitting electrons, when we have a spin-flip, then the boundary term reads: 
$$H_{boun}^{s-flip}(x=0)\propto \mathcal{T}_0 \cos \theta_{\rho \pm} \cos \phi_{\sigma \pm}$$
and when we have a valley-flip, then the boundary term reads:
$$H_{boun}^{v-flip}(x=0)\propto \mathcal{T}_0 \cos \theta_{\rho+} \cos \phi_{\rho-}  \cos \theta_{\sigma+} \cos \phi_{\sigma-}$$
By comparing these last equations with Eq.\ref{eq:standard-with-sf} and Eq.\ref{eq:standard-with-sv}, respectively, we observe that when switching between the usual (pinning) and the radiative BC we switch between canonically conjugated $\vartheta_{\nu}$ fields in the general expression Eq.\ref{eq:ham-bound}. In this work we are not focusing on these spin-flip and valley-flip processes; however, since such boundary conditions are possible, we include them here for the sake of completeness. It has to be emphasized that all these cases can be accommodated in our formalism simply by changing the sign of $b_{c,s}$, the exponents of spatial dependence. To be more precise: if we have TLL with $K_\nu\approx 1/4$ then we set radiative BC for the $\cos\theta_{\nu}$, while if $K_\nu\approx 4$ then we ought to set the pinning BC for the $\cos\phi_{\nu}$ such that in either case $b_{\nu}\approx 2$ and our analytical formulas are applicable.

\paragraph{Example of a nanotube (i.e. two leg ladder)} Let us consider a quantum dot attached on the top of a nanotube with intense co-tunneling processes through the electron-emitting dot. Then we shall have radiative boundary condition for $\rho+$ which will always have $\theta_{\rho+}$ frozen, thus the boundary exponent will depend on $K_{\rho+}$. If the quantum dot is large (few atoms or more) then the largest interaction (from Coulomb blockade) is between electrons of the same spin-valley and plain radiative condition, Eq. \ref{eq:ourBC} applies. The boundary exponent will depend on $\sum_{\nu}K_{\nu}^2$ which is already the unusual case. There are however further complications possible.  For a single atomic impurity Hubbard-type term implies the presence of the spin- and valley-flip processes. Thus, the radiative boundary condition will then be set for $\phi_{\sigma+}$ or $\phi_{\rho-}$. These will be the fields frozen on the boundary. Therefore, as usual in 1D boundary, the boundary exponent will depend on $K_{\rho+}^2+1/K_{\sigma+}^2$ or $K_{\rho+}^2+1/K_{\rho-}^2$. 

To complete the discussion we look at backscattering terms: if we consider a magnetic ad-atom on the top of nanotube without emission then there will be standard $\cos\phi_{\rho+}$ boundary condition for the holon, but $\cos\theta_{\sigma+}$ for the spinon mode, which will now lead to boundary exponent that depends on $K_{\sigma+}^2+1/K_{\rho+}^2$. 


\section{Results}

\subsection{Analytic expression}

Physically, accessible quantities are usually measured in the frequency domain. Consider, for instance, an STM measurement done along a 1D wire. 
The expression, as given above, thus needs to be partially Fourier transformed; namely, we need to perform the integral transform $\mathfrak{F}[]$, but only along the time-axis, into the frequency domain, the $\mathfrak{F}_{t\rightarrow\omega}[N(r,t;T)]$. Although it sounds rather straightforward, this is in fact a big technical issue. Already for the simplest case of a single-mode TLL, the Fourier transform of the finite temperature expression takes the form of hypergeometric Beta function [\cite{giamarchi2003quantum}]. Analytical formulas are known only for the cases with two velocities (as Appel hypergeometric function
\cite{iucci2007fourier}), and for the boundary problem, the formula is expected to be even more complicated. On the other hand, analytical formulae have a great advantage, since both the $N(x\rightarrow 0,r,t)$ and $N(r,\omega)$ have singularities, thus numerical integrals are notoriously hard to control. This is particularly important if we with to use $N(r,\omega)$ as an input for some further calculations.  

We thus see that obtaining $N(r,\omega)$ is a non-trivial but important task that so far has been done only for a few special cases. We decided to search for special values of $b_i$ where the transform is possible. Previously\cite{grigoryan2024role} we obtained the result for the case $b_s=0$ and $b_c=2$, here we show a new expression for the case $b_s=2$ and $b_c=2$. When we have a one leg-ladder and the two $b_\nu$ parameters are fixed, then we have fixed both $a_\nu$ parameters, while for $\bar{N}$ leg-ladder, we can still vary the  $a_i$ parameters, which means we can vary the $K_\nu$ parameters. Both expressions are given explicitly, in their full form in App.B.

Although the formulas themselves are quite lengthy, some useful remarks can be drawn from their functional form. We see that each formula consists of several building blocks with hypergeometric $_2F_1()$ function in the numerator. Some of these come with an extra $\pi$ phase shift in the argument and with spatially dependent pre-factors $\sim v_{\rho} r$. For Hund metal case, there are two types of spatially dependent pre-factors $\sim (v_{\rho}+v_{\sigma})/2 r$ and $\sim (v_{\rho}-v_{\sigma})/2 r$, which is a manifestation of the fact that i) now both spin and charge contribute to spatial dependence; ii) the two waves can interfere. One can use the known relation between the hypergeometric $_2F_1$ function and the incomplete Beta function:


\begin{equation}
    B_{\varkappa}(\bar{a},\bar{b})=\frac{\varkappa^{\bar{a}}(1-\varkappa)^{\bar{b}-1}}{\bar{a}}\,_2F_1\left(1, 1-\bar{b}, \bar{a}+1; \frac{\varkappa}{\varkappa-1}\right)
\end{equation}

which upon substitutions

$$\varkappa \rightarrow \left(1-\exp{\left[-\frac{2\pi}{\beta}t \right]}\right)^{-1}$$

$$\bar{a}\rightarrow \frac{1}{2}\left(a_c+a_s+\frac{\imath\beta\omega}{\pi}\right)$$

$$\bar{b}\rightarrow \frac{1}{2}\left(a_c+a_s-\frac{\imath\beta\omega}{\pi}\right)$$

proves that our formula can be re-expressed in terms of incomplete hypergeometric Beta functions, thus making a connection with well-established bulk TLL results: the Fourier transform in the bulk TLL is given in terms of the Beta function\cite{giamarchi2003quantum}. Here, we show that the presence of boundary and the second TLL mode generalizes that expression into a combination of incomplete Beta functions, allowing for a connection between the two situations. We find here the \emph{incomplete} Beta functions, just like in Ref. [\onlinecite{chudzinski2020contribution}], where it was shown that this corresponds to a possibility of a finite diffusion time in the open system. We elaborate more on this at the end of App.B.

The advantage of writing the expression in a less compact form, with $\bar{a}=\omega+K$ in the denominator, is that it manifestly has the form of a Lehman representation for a free boson propagator. We computed the local propagator, a quantity that is integrated over all momenta, to obtain the above-mentioned denominator ($\pm \pi$) which implies that original LDOS has a plasmon pole form with bosons moving along the light rays defined as Dirac deltas $\delta(k-K_{\nu}/\beta)$. In our procedure to obtain correlation functions, we have re-scaled fields by $K_{\nu}$ to arrive at the non-interacting theory, without Bogoliubov angle that would need to be non-local when left- and right-going fermions are coupled. In this case, as it has been recently proven, the $K_{\nu}$ are becoming related to Thomas-Fermi screening length. Thus the shift of the plasmon pole can be interpreted as a characteristic screening length in the material. 

Finally, we decided to keep the formula in its most general form as it will later enable us to generalize to complex TLL parameters that will encode non-Hermiticity in the presence of a strong external field.

While this is a result obtained at a particular point, it should be emphasized that since the $N(r,\omega;\beta)$ is a continuous function of $K_{\nu}$, we do expect that our results give a good indicator for overall behaviour of this function in a strongly correlated case. Furthermore, the obtained value lies close to a limiting (separatrix) behaviour of a sine-Gordon model, i.e., a TLL with a cosine perturbation where the latter term may be due to backscattering. Thus, we expect that this result will arise frequently in real-life experiments. Finally, it should be noted that although we solved the problem with two bosonic poles (velocities), this can correspond to the situation with indeed only two modes, as well as to the situation with more modes provided, there are only two velocities (i.e., some modes are degenerated).

The $b_s=0$ case corresponds to the situation when $K_s=1$ (all $K_{\nu}=1$ for all neutral modes), which is a non-interacting value. This can be realized for purely charge and long-range interactions such as Coulomb interactions. The $b_s=b_c$  case implies that both spin and charge sectors are equally shifted away from the non-interacting point. This can be realized for on-site interactions under the condition that only parallel spin electrons interact. Such a situation may arise, for instance, in a Hund metal, where $J_H$ dominates low-energy physics. 

\subsection{LDOS $N(r,\omega;\beta)$}

Here, we compare the results of integral transforms obtained for Coulomb and Hund metals. We plot them as a function of a distance from the boundary $r$ and energy away from chemical potential $\omega$. There are several panels, each at a different inverse temperature. As a function of a distance $r$, we observe that when we move away from the TLL's boundary, there is a sudden drop of LDOS close to the boundary in all cases, followed by a broad maximum (in Hund metal very shallow) and a gradual decrease towards bulk behaviour.





Fig. \ref{fig: DOS_columb} demonstrates TLL DOS for Coulomb metal as a function of energy $\omega$ and the distance from the boundary  $r$, where $\omega$ is measured relative to the Fermi energy. Here, essentially everything shows the behavior of a power law as a function of frequency at both small and large values of $r$. 
An interesting feature is that at $\omega = 0$,  the energy corresponds to the Fermi level, and a phenomenon of a pseudo-gap is observed.
This is described in more detail in  Ref. [\onlinecite{grigoryan2024role}].

The appearance of the broad peak is somewhat expected. From Eggert, we know that LDOS should be a function of a variable $x\omega$. For the $\omega$ dependence of LDOS, we know that it scales as $\alpha_{bulk}-1$ (in the bulk) or $\alpha_{edge}-1$ (on the boundary). While $\alpha_{bulk}$ is always greater than one, upon choosing radiative boundary condition with $K_{\sigma}<1/2$, we took $\alpha_{edge}<1$ which gives an inverse scaling of LDOS close to the boundary.

\begin{widetext}

\begin{figure}
    \centering
    \includegraphics[width=0.4\textwidth]{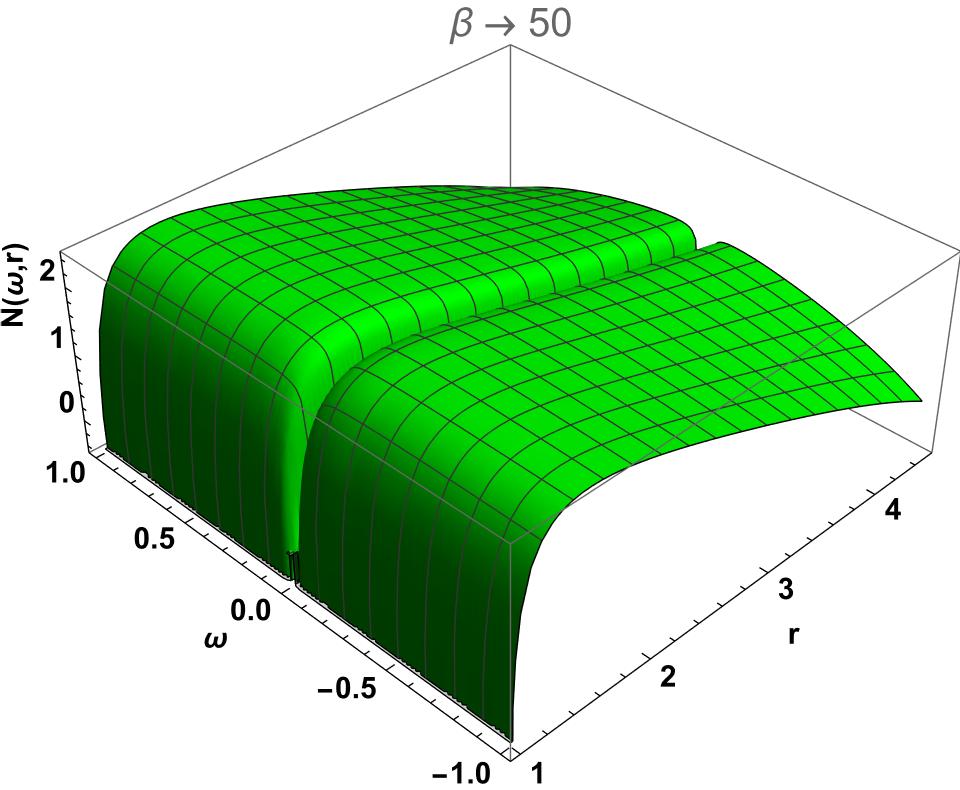} \hspace{1 cm}
    \includegraphics[width=0.4\textwidth]{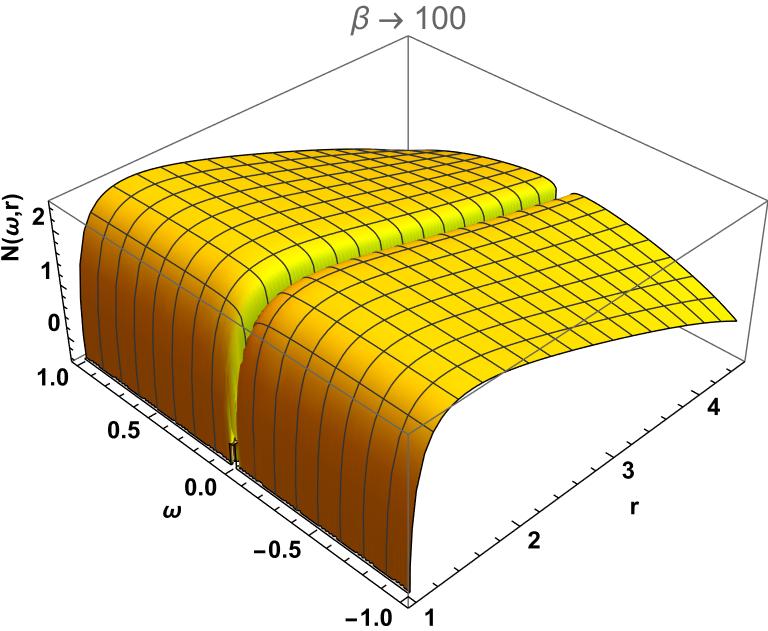} \\[0.3 cm]
    (a) \hspace{6.5 cm} (b)\\[0.3 cm]
    \caption{TLL DOS for \textbf{Coulomb metal} as a function of energy $\omega$ and the distance from the boundary  $r$ (tip of the carbon nanotube, unit of $r$ is $1/V_F$); when (a) $\beta = 50$, (b) $\beta = 100$. 
    The $\beta=1/T$ is the inverse temperature with a unit set by the fact that the unit of energy was set by $v_F=1$.}
    \label{fig: DOS_columb}
\end{figure}


\begin{figure}
    \centering
    \includegraphics[width=0.4\textwidth]{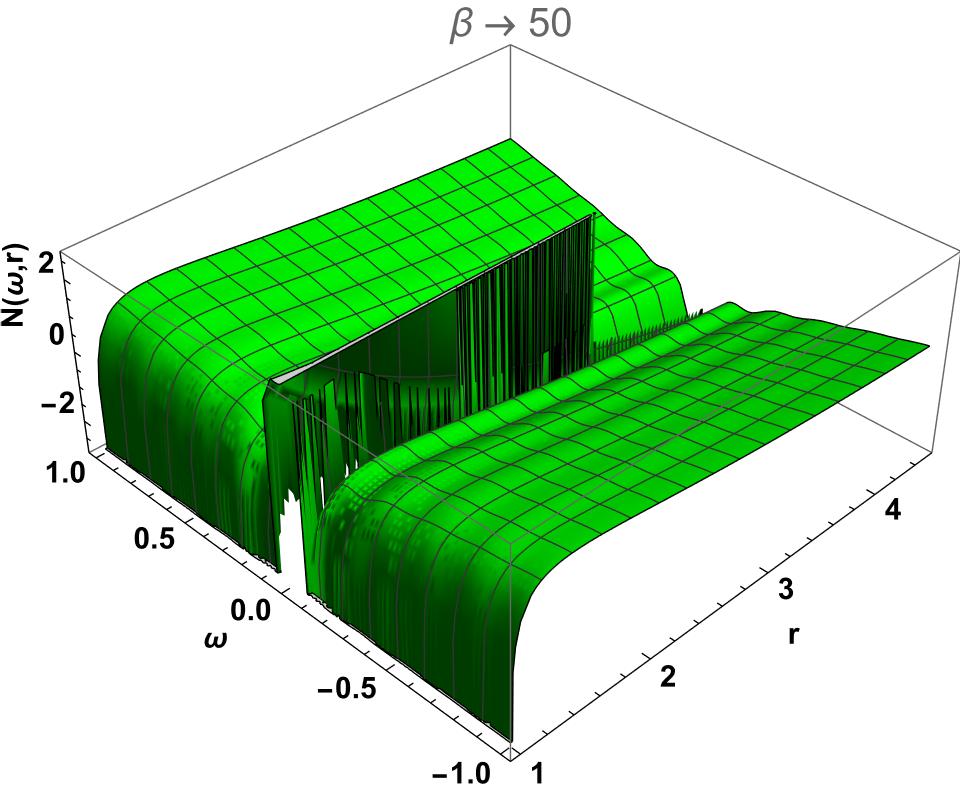} \hspace{1 cm}
    \includegraphics[width=0.4\textwidth]{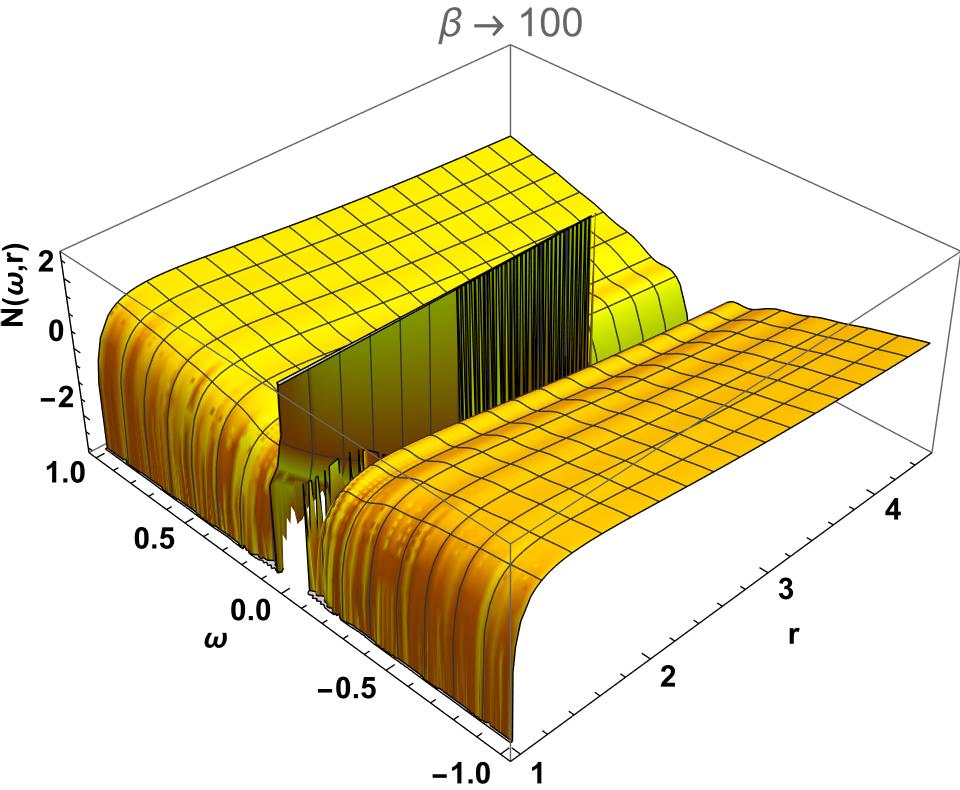} \\[0.3 cm]
    (a) \hspace{6.5 cm} (b)\\[0.3 cm]
    \caption{TLL DOS for \textbf{Hund metal} as a function of energy $\omega$ and the distance from the boundary  $r$ (tip of the carbon nanotube, the unit of $r$ is $1/V_F$); when (a) $\beta = 50$, (b) $\beta = 100$. 
    The $\beta=1/T$ is the inverse temperature with a unit set by the fact that the unit of energy was set by $v_F=1$.}
    \label{fig: DOS_Hunds1}
\end{figure}

\begin{figure}
    \centering
    \includegraphics[width=0.4\textwidth]{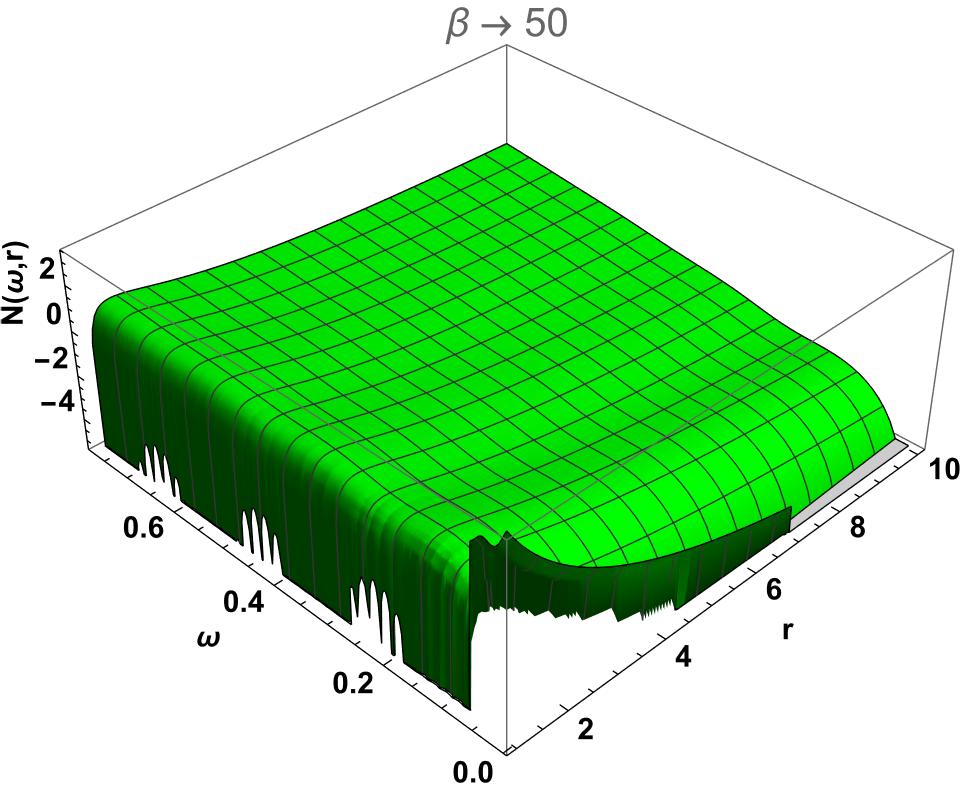} \hspace{1 cm}
    \includegraphics[width=0.4\textwidth]{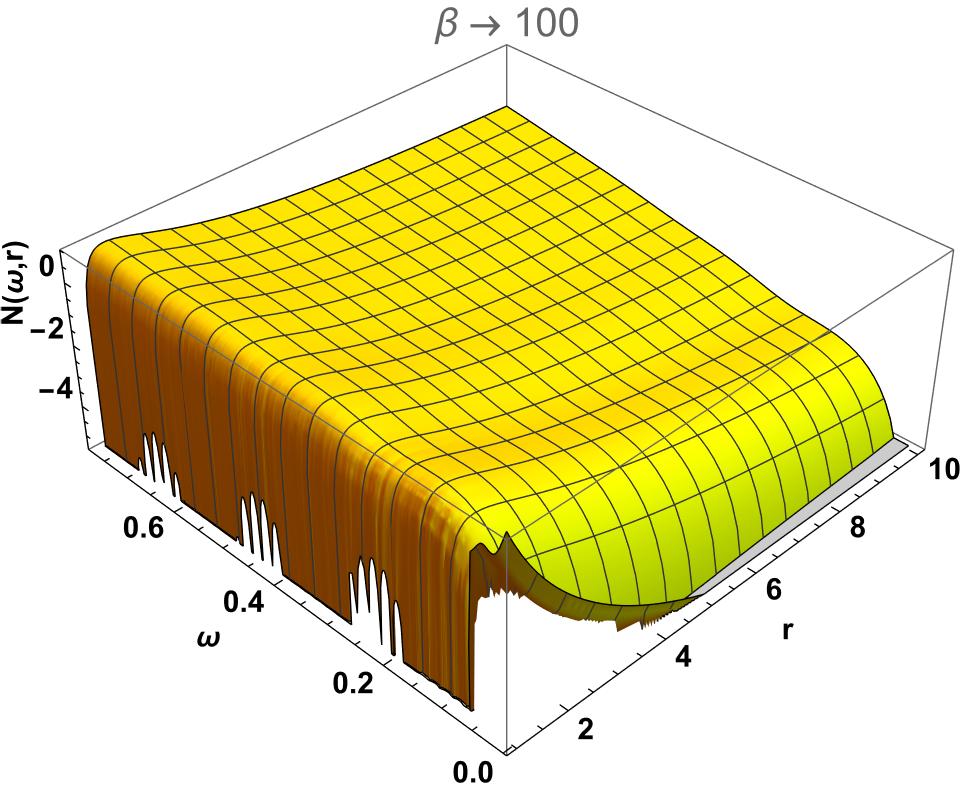} \\[0.3 cm]
    (a) \hspace{6.5 cm} (b)\\[0.3 cm]
    \includegraphics[width=0.4\textwidth]{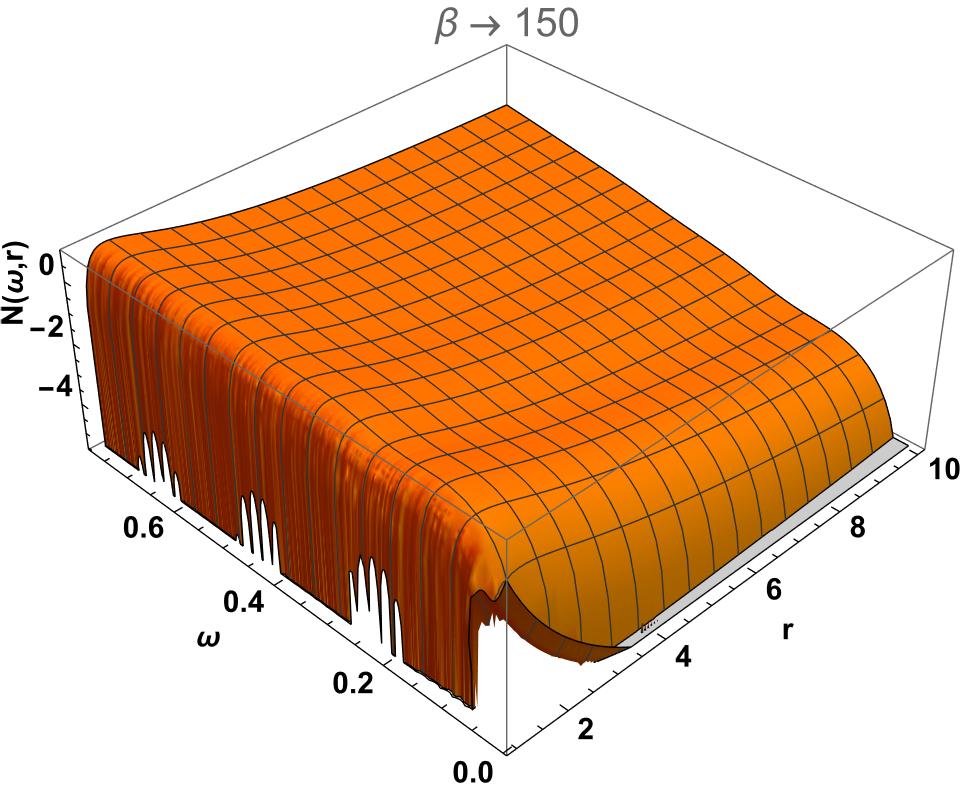} \hspace{1 cm}
    \includegraphics[width=0.4\textwidth]{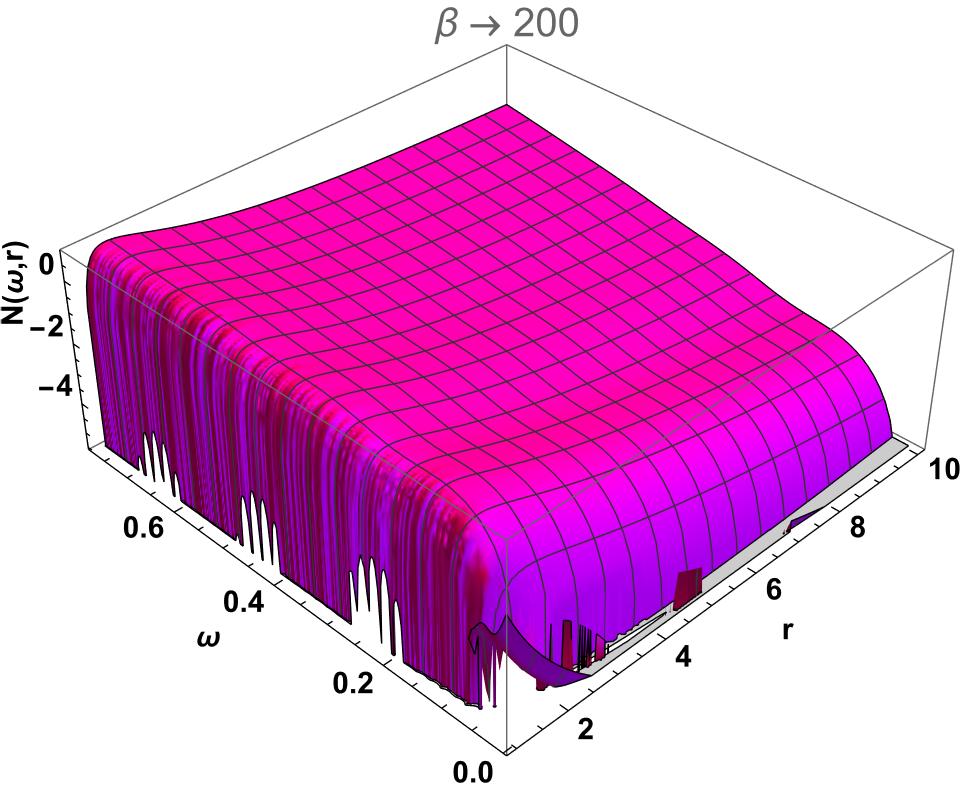}\\[0.3 cm]
    (c) \hspace{6.5 cm} (d)\\[0.3 cm]
    \caption{TLL DOS for \textbf{Hund metal} as a function of energy $\omega$ and the distance from the boundary  $r$ (tip of the carbon nanotube, unit of $r$ is $1/V_F$); when (a) $\beta = 50$, (b) $\beta = 100$, (c) $\beta = 150$, (d) $\beta = 200$.  The $\beta=1/T$ is the inverse temperature with a unit set by the fact that the unit of energy was set by $v_F=1$.}
    \label{fig: DOS_Hunds}
\end{figure}

\end{widetext}
We see that the same LDOS can be produced for $K_{\nu}$ and $1/K_\nu$, provided the boundary conditions are reversed. For a more intuitive $\phi_\nu(x=0)$ boundary condition, we have strongly attractive interactions in a charge sector $K_c\gg 1$. For the Coulomb metal case, this results in a broad maximum in the vicinity of the boundary which suggests that electrons accumulate close to the boundary because it is energetically favorable for them to stay together in a constrained area. Similarly, the $K_s\gg 1$ for the spin sector implies that we are in an Ising ferromagnet regime, where the system has a tendency to develop easy-axis large local magnetization that decouples from the boundary at the lowest temperatures. 

The situation for strong repulsive interactions $K\ll 1$, which is frequently more physically relevant, is even more interesting. We see that the radiative boundary condition $\theta_c(x=0)$ for the charge sector induces an enlarged LDOS close to the end of the 1D system, which suggests that charge accumulates there to facilitate in-out charge fluctuations. 
The fact that in Hund metal both spin and charge modes with different velocities contribute to the spatial dependence of LDOS, also manifests as its weak oscillations, the beating phenomena which are visible most prominently for the smallest $r$.

An interesting dichotomy can be observed at the lowest energies. In the case of Coulomb metal (Fig. \ref{fig: DOS_columb}), the LDOS has a simple zero bias anomaly, an interaction-induced pseudo-gap, as the Fermi level is approached. On the contrary, the most remarkable feature of Hund metal (Fig. \ref{fig: DOS_Hunds1}) is the presence of a sharp peak in close vicinity of $\omega=0$, which decays with $r$ relatively quickly. This can be interpreted as an interference phenomenon. In the Coulomb metal case, not only is there no interference, but also any feature close to zero energy is suppressed by boundary condition, affecting phase shifts (that mathematically enter through partial integration explained in App.A). Curiously, the peak seems to disappear as we move towards the lowest temperature. Initially, the numerical accuracy was blamed for that (as the standard Kondo peak is expected to narrow), but extensive trials of different meshes were not able to change this picture. It is thus necessary to understand better the specific regime of the spin system being tackled here.



This feature of Hund metal is illustrated in more detail in Fig. \ref{fig: DOS_Hunds} which represents the same features as in Fig. \ref{fig: DOS_Hunds1}. We used a smaller energy range and several temperatures, to illustrate more effectively what is going on close to the $\omega = 0$. It depicts the temperature effect:  at higher temperatures, the peak is stronger; at lower temperatures, it becomes weaker.

\subsection{Interpretation of the $\omega \rightarrow 0$ features.}

The situation is quite remarkable in the Hund metal case, where both modes are deeply in the relevant regime: here we detect a strong, sharp peak close to the zero energy. In order to understand this feature let us discuss the spin sector where the interpretation is more transparent. In our range of parameters $K_s\ll 1$, the system would be deep inside the antiferromagnet (AFM) Ising phase, however, in the absence of cosines gapping term in the bulk Hamiltonian Eq.\ref{eq:ham-TLL}, it remains gapless. We thus expect to work in the TLL phase but in its strongly anisotropic $J_z\gg J_{\perp}$ limit. Adding a boundary cosine (Eq. \ref{eq:ham-bound}) to the TLL Hamiltonian (Eq. \ref{eq:ham-TLL}) implies that the model \emph{locally} turns into the sine-Gordon model. The "natural" excitations of this last model are solitons and possibly their combined states: breathers. For the $\cos\theta_{\nu}$ perturbation the breathers would be allowed only for large enough values of $K_{\nu}$ which is far away from our regime. 

Formally, we have a real space expression for the correlation function $N(r,t;T)$ and we need to understand how the peak in $N(r,\omega;T)$ can emerge from it. To rationalize it we write down a formal expression for the chiral Green's function $G_{RR}(r,\omega;T)$, since the desired $N(r,\omega;T)\propto Im[G(r,\omega;T)]$:
\begin{multline}
  G_{RR}(r,\omega;T)=\langle \psi_R^{\dag}(r,\omega)\psi_R(r,\omega)\rangle\\
  = \int d\psi_R \psi_R^{\dag}(r,\omega)\psi_R(r,\omega)\exp(S_{TLL}+S_{cos})   
\end{multline}
where the crucial point is that one takes \emph{both} the TLL and the cosine part of the action. It is then assumed that the correlation function can be computed with $S_{TLL}$ only provided the cosine has been absorbed in step 3 of \textbf{BP}. One thus uses the non-interacting bosonic fields $\tilde{\phi}_R$ defined in step 2 of \textbf{BP} computes the TLL power laws from $
\exp\langle\phi_R\phi_R\rangle_{S_{TLL}}
$ implicitly assuming that:
$$
\int d\psi_R \psi_R^{\dag}(r,\omega)\psi_R(r,\omega)\exp(S_{cos})=0
$$
simply because the cosine has already been absorbed. Let us now take a closer look at that. By taking a Taylor expansion of the exponential $\exp(S_{cos})=1+V_0\prod_{\nu} \cos\theta_\nu(x=0)+...$ we see that the first term indeed leads to the TLL power-law, but the latter term gives a cosine of the original $\theta_{\nu}(x=0)$ field. We can now combine the exponentials of the bosonic fields to arrive at:
$$
\int d\tilde{\phi}_{\nu}d\tilde{\theta}_\nu \exp(i\sum_{\nu}\tilde{\phi}_\nu+\tilde{\theta}_\nu+\theta_{\nu})|_{r=0,\omega}\exp(S_{TLL})
$$
\begin{widetext}
keeping in mind step 2 of \textbf{BP}, the relation $\theta\rightarrow\tilde{\theta}$, this can be rewritten as:
$$
\int d\tilde{\phi}_{\nu}d\tilde{\theta}_\nu \exp(i\sum_{\nu}\sqrt{1+K_{\nu}}[1/\sqrt{1+K_{\nu}}\tilde{\phi}_\nu+\sqrt{1+K_{\nu}}\tilde{\theta}_\nu])|_{r=0,\omega}\exp(S_{TLL})
$$
now, if and only if $K_{\nu}$ for all modes are equal we can take the $\sqrt{1+K}$ outside the sum and perform a unitary transformation $U=\exp(\sqrt{1+K}\sum_{\nu}\theta_{\nu})$ which leaves us with a definition of a new spin-full fermion:

\begin{equation}\label{eq:referm}
   \psi_0=\exp(i\sum_{\nu}[1/\sqrt{1+K_{\nu}}\tilde{\phi}_\nu+\sqrt{1+K_{\nu}}\tilde{\theta}_\nu]) 
\end{equation}
and we are now solving an instanton-type problem with boundary scattering since the unitary transformation has added a boundary state to the bulk $S_{TLL}$ action.

\end{widetext}

\begin{figure}
    \centering
    \includegraphics[width=0.45\textwidth]{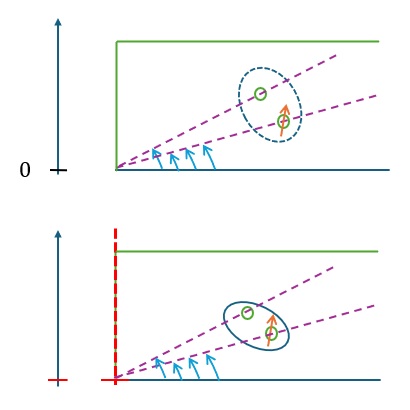}
    \caption{Top: Spinon and holon bound state is transient.
    Bottom: Fermion hybridization in a TLL, when $K(r) = cste$.
    The green lines represent the Luttinger parameters, the dashed purple lines correspond to bosonic dispersions, and the blue arrows indicate thermal excitations. }
    \label{fig:K-HundvsCoul}
\end{figure}

We thus approach the single-particle wave scattering regime for which we know what is the outcome of the boundary problem. To be more precise, due to our boundary condition set on the $\theta$ field, we are picking the odd solutions of the 1D barrier problem. For a strictly local, $x=0$, boundary, there exists only one solution in this class, the Dirac-delta $\psi_0$ with $E_{0}=0$. This is illustrated as the red $\delta$ state in Fig.\ref{fig:K-HundvsCoul}. It should be emphasized that for the standard $\cos{\phi_{\nu}(x=0)}$ boundary potential, the even boundary condition, there is no additional boundary state, thus there is indeed no extra boundary LDOS and one observes pure pseudo-gap (also known as the zero bias anomaly).

From the perspective of the remaining part of TLL, the bath, the radiative boundary condition represents strong Kondo-type coupling with an outside entity, and this is a spin-flip type coupling $\sim \exp(i\theta_s)$.  These particles will contribute to the spin-flip processes described above, the processes that are usually captured by means of re-fermionization into some emergent $\psi_0$ state.
We can now understand the difference between Coulomb and Hund metal: what manifests itself as a phase shift on a mathematical level (see App. A for details), on a physical level an equal value of compressibilities for spin-and charge sectors $K_{\rho}=K_{\sigma}$ allows for a re-fermionization of a full spin-full fermion $ \tilde{\psi}_R$. This is the particle that is visible in our plots of LDOS. 
This idea is illustrated in Fig. \ref{fig:K-HundvsCoul} where the energy spectrum and hybridization processes in a TLL system with constant and spatially varying Luttinger parameters $K(r)$ are shown and compared for Coulomb (top) and Hund (bottom) cases.

In the top panel, $K$ parameters are different for spinon and holon, and therefore, it is not possible to write the Eq. \ref{eq:referm}. 
In the bottom panel $K(r)=const$, a quantum well forms near the boundary, resulting in a hybridization. Here, a particle is created inside the TLL, where we combine a holon and a spinon to form a fermion-like object which can jump back and forth, resulting in an extra state.

The observed temperature dependence (Fig. \ref{fig: DOS_Hunds}) is in agreement with this picture. At the lowest temperatures i.e. close to the fixed point where $\sim\cos\phi_\nu(x)$ would be relevant (if it existed), bulk and boundary are incompatible, in the sense that an easy axis bulk that favours $\sim \phi_{\nu}$ locking cannot provide so much radiation coupling $\sim\cos\theta_s(x=0)$ with an in-plane object oscillating outside the TLL. For this the difference of excitations ($q\neq 0$) $\frac{1}{K_{\nu}}(b^{\dag}_{q}-b_{-q}^{\dag})$ matters, especially for small $K_{\nu}$. The situation changes at finite temperatures when the TLL bosons $b_{q}$ start to be excited and they can couple to the boundary state. Thus we observe a peak with an amplitude that increases with the temperature.



\section{$4k_F$ charge susceptibility}

A similar approach to spectral function can be applied to two-body correlations, the susceptibilities. Of special interest is the charge susceptibility as it determines screening, and thus, the local dielectric constant. In our range of TLL parameters, when $K_{\rho}<1/3$, the slowest decaying correlation is the $4k_F$ susceptibility. It represents a combination of two $2k_F$ susceptibilities; while in the bulk, it is $\propto \chi_{2kF}(x,t)\otimes \chi_{2kF}(x,t)$, on the boundary, there is also a term $\propto \chi_{2kF}(r,x,t)\otimes \chi_{-2kF}(-r,x,t)$ which shall contribute to the uniform component. The $\chi_{4kF}$ depends on $\cos(4\phi_{\rho})$ operator, without a component from the spin sector. Since the difference between Coulomb and Hund metals is in the spin sector, this dominant contribution to charge susceptibility is going to be the same both for Coulomb and Hund metals.

\begin{widetext}
\begin{multline}
    \chi^{TLL}(\omega,r) = \int dt \cos (\omega (t) \left(\frac{1}{\sinh ^2\left(\frac{\pi  t}{\beta }\right)}\right)^{2 K_c}
    \left(\frac{1}{\sinh \left(\frac{\pi }{v_c \beta} \left(2 r + v_c t\right) \right) \sinh \left(\frac{\pi }{v_c \beta}  \left(2 r - v_c t\right)\right)}\right)^{2 K_c} \\
\end{multline}

For the case when $\omega=0$ and $2 K_c = 1$, which we are particularly interested in when calculating static interactions that determine $K_{\rho}$, the above integral can be evaluated analytically. The result reads:

\begin{multline}
    \chi^{TLL} = \frac{1}{4 \pi} \beta  \text{csch}\left( \frac{2 \pi r}{\beta v_c} \right)^3 \text{sech}\left( \frac{2 \pi r}{\beta v_c} \right) \left( -2 \log \left( \sinh \left( \frac{\pi ( -2 r + t v_c)}{\beta v_c} \right) \right) \right. \\
    \left. + 2 \log \left( \sinh \left( \frac{\pi ( 2 r + t v_c)}{\beta v_c} \right) \right) + \text{csch}\left( \frac{\pi t}{\beta} \right) \left( \sinh \left( \frac{\pi ( -4 r + t v_c)}{\beta v_c} \right) - \sinh \left( \frac{\pi ( 4 r + t v_c)}{\beta v_c} \right) \right) \right)
\end{multline}
\end{widetext}

\begin{figure}
    \centering
    \includegraphics[width=0.45\textwidth]{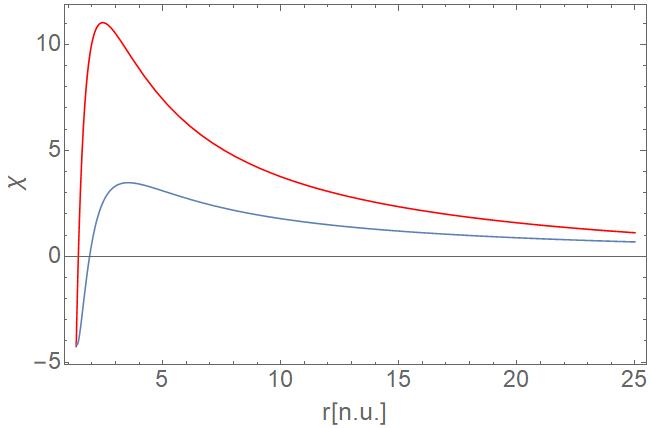}
    \caption{Illustration of the static, $\omega = 0$, $q=4k_F$ real part of charge susceptibility $\chi(r;\omega=0,q=4k_F,\beta)$ as a function of a distance from the boundary for two temperatures $\beta=50$ (red) and $\beta=100$ (blue).}
    \label{fig: charge susceptibility}
\end{figure}

This result can be regularized in a similar manner like the LDOS. The resulting susceptibility is plotted in Fig. \ref{fig: charge susceptibility}. As before, we see a cross-over between two power laws: a boundary and a bulk, with the boundary increase much steeper and the bulk decay much slower. This is in agreement with the result for longitudinal spin susceptibility reported in 
\cite{schneider2022boundary} if one recalls that we work in the limit of small $K_{\rho}$. The TLL temperature dependence is due to substitution $r^{-2a}\rightarrow \beta^{-2a}Sinh^{-2a}(r/\beta)$, and indeed, the two curves converge for small $r$ while at finite $r$, we observe larger signal for smaller $\beta$ and the shift from maximum to larger values with increasing $\beta$.  Remarkably, as the temperature increases, the boundary peak of a two-body charge susceptibility is much more prominent than the single-particle boundary feature in the Coulomb metal, which suggests that probes measuring local susceptibilities like relaxation-time $1/T_1$ NMR or Raman spectroscopy may be better suited to study these many-body changes of profile than more frequently used STM.

Based on the RPA, which is known to hold in TLL, one can now estimate the dielectric function $\epsilon(r)=1+ c_{back} N_{em}\chi_{\rho}(r)$, where $N_{em}$ is the density of emitted carriers and $c_{back}$ is a geometry-dependent constant that describes the strength of the back-reaction onto the TLL. At this point, it should be emphasized that from the Dzyaloshinskii-Larkin theorem, vertex corrections are prohibited in a pure TLL. Thus, any correction to the dielectric function and therefore to the electron-electron interactions is possible only due to the presence of emitted carriers. In the zeroth order approximation, only those outside carriers are going to be affected by the spatially dependent screening, the $\epsilon(r)$. When the density of emitted carriers increases, then this decoupling approximation will break down; this regime will be explored in the next section.  

On the bulk side (large $r$ values), it can be observed that the dielectric function from TLL carriers initially increases as one approaches the end of the 1D system. This has important consequences, for instance, in the case of field emission. From electrostatics, it is expected that inter-tube screening should decrease and the external electric field should increase as the emitted carrier approaches the tip of the 1D system. However, the increase $\epsilon(r)$ when $r\rightarrow 0$ as observed here for $r>1[n.u.]$, implies that this electrostatic effect is at least partially compensated by correlations, thus, for instance, an assumption of constant screening along the 1D system, which is used in field emission studies, is justified. 

On the boundary side, a rapid increase in charge susceptibility is observed. This implies that somewhere close to the boundary, the relative dielectric constant will be smaller than one, and this will be a manifestation of strong radiation set on our boundary. In principle, this should be the point where the boundary condition is set, as a further decrease towards a negative relative dielectric constant indicates that our formalism would be breaking down therein. Here, employing the strong coupling approach becomes necessary.

\section{Spatially dependent TLL parameters}

In the previous section, we have considered a weak coupling regime of emission, namely a situation when the cloud of emitted carriers has no feedback effect on the TLL emitter, their density $N_{emit}\rightarrow 0$. In this section, we would like to move beyond this regime. In the strong coupling limit, one can expect the presence of plasmon-polariton quasi-particles that will couple with holons (1D plasmons). There will be two new effects that need to be considered: \textbf{i)} the external carriers will modify screening inside TLL; \textbf{ii)} the carriers from TLL will randomly jump onto the polariton state and back.

We first focus on interaction effect \textbf{i)} that modifies the TLL parameters and velocities, thus can be considered a correction to the real part of bosonic self-energy. The TLL parameters are determined by electron-electron interactions. In particular, the long-range interactions determine the value of charge mode $K_{\rho}$. As stated above, due to the Dzyaloshinskii-Larkin theorem, the TLL cannot screen itself, but carriers surrounding it will be affected. For a sufficiently large $N_{emit}$, it will be more appropriate to consider two coupled metals, the TLL and the plasmon-polariton cloud, that are hybridized together since the charge can flow in between them. Following the Maxwell-Garnett equation, from the effective medium approximation for nano-rods 
\cite{resano2015experimental}, one finds:
\begin{equation}
   \epsilon_{eff}(r)=\frac{(1-f)\epsilon_{TLL}(r)+f\beta_{dep}\epsilon_{pp}(r)}{1-f+\beta_{dep}f} 
\end{equation}
where $f=N_{emit}/N_{TLL}$ and $\beta_{dep}\sim O(1)$ is a constant -- a depolarization factor,  e.g. $\beta_{dep}=1/2$ for a rod\cite{JONES2003284}. We deduce that when a density of carriers inside TLL is much larger than $N_{emit}$, which is still a sensible assumption, then the TLL charge susceptibility illustrated in Fig. \ref{fig: charge susceptibility} will become the charge susceptibility of the entire medium. To be precise, in the RPA, the dielectric function is directly proportional to susceptibility $\epsilon_{TLL}(q;r)=1-\tilde{V}_{Coul}\chi_{TLL}(q;r)$ where we took the static limit $\omega\rightarrow 0$.  

The charge TLL parameter depends on screened Coulomb interactions in the low momentum exchange limit:
$$K_{\rho}(r)=1/\sqrt{1+\tilde{V}_{Coul}(q\rightarrow0,r)}$$
where 
$$\tilde{V}_{Coul}(q\rightarrow0,r)=V_{Coul}(q\rightarrow0)/\epsilon(q\rightarrow0,r).$$
Once we know how dielectric constant $\epsilon_{eff}(q\rightarrow0,r)$ and thus screened interactions $\tilde{V}_{Coul}(q\rightarrow0,r)$ depend on $r$, we can deduce the dependence $K_{\rho}(r)$. Then, in principle, we can incorporate this effect together with velocity variation $v_{\rho}(r)$ in the LDOS profile. However, to profit from our analytical solution for Fourier transform, we still need to fulfil the condition $b_c=2$, which apparently cannot be done any more because the $|K_{\rho}|$ has changed.

Fortunately, our modelling cannot stop at this point as we still need to take into account electron jumps between the TLL and the plasmon-polariton cloud. This is the second \textbf{ii)} effect indicated at the beginning of this section. These incoherent jumps were the necessary ingredient that enabled us to use the effective medium approach. In a recent paper 
\cite{yamamoto2022universal}, it has been shown how to treat the TLL subjected to incoherent coupling with an external level $b_0$; a problem has been treated using Lindbladian formalism with the following Lidblad master equation:
\begin{equation}
    \frac{d\rho}{dt}=\imath[H_{TLL},\rho(t)]+ \hat{L}^{\dag}\rho\hat{L}
\end{equation}
where $\rho$ is a density of states of the system under consideration and the jump operator is:
\begin{equation}\label{eq:Lindb-def}
  \hat{L}(x)= \sqrt{\gamma(x)}b^{\dag}_{\rho+}(x)b_{\rho+}(x)  
\end{equation} 
where $b^{\dag}_{\rho+}$, $b_{\rho+}$ are the bosonic creation/annihilation operators corresponding to the field $\phi_{\rho+}$, and since plasmon-polariton coupling is originally due to Coulomb-like force, we expect it to interact only with charge-full particles (in other words, the interaction cannot distinguish spin or valley degree of freedom). The $\gamma(x)$ is the amplitude of bi-linear coupling with an averaged plasmon-polariton liquid $\gamma(x)\propto d_{pl}(x=r)=\langle b_0^{\dag}(r)b_0(r)\rangle$. Such coupling between TLL and bosons with linear dispersion has been accounted for in Ref. [\onlinecite{Martin-Loss-WB}].

In the rotating frame (secular) approximation, the Lindbladian superoperator reduces to the Redfield equation, and then the following effective Hamiltonian describes the quasi-open system:
\begin{equation}
  H_{eff}= H_{TLL} - \frac{\imath}{2}\int dx L^{\dag}(x)L(x)  
\end{equation}
the shift depends on local strength of Coulomb interaction $\gamma(x)=\gamma_0 V_{loc}(x)$ with $\gamma_0\propto N_{emit}$. Following Ref. [\onlinecite{Martin-Loss-WB}], we know that the coupling is of a displacement type: hence, $V_{loc}(x)\propto \nabla_x$. Moreover, those authors showed that integrating out the $d_{pl}$ boson bath leads to TLL with modified parameters, thus here we also expect an additional (imaginary) contribution to the TLL parameter $K_{\rho}$.

We can now investigate how the procedure \textbf{BP} needs to be modified when $K\rightarrow K+\imath\gamma(k)$. In the above, we have written $\gamma(k)\propto k$, which is a Fourier transform of $\gamma(x)\propto \nabla_x$. One notes that since the coupling with bath, the imaginary part of $K$, is proportional to momentum, thus it changes sign when moving from right to left going fermion. Again, the delicate step is imposing boundary condition, the step \textbf{3} of \textbf{BP}. Thus, now we shall have the relation:
\begin{equation}
    \phi_L(x,K)= - \phi_{R}(-x;K^*)
\end{equation}
where we indicated canonically conjugated $K^*$. As a result, the non-local term $b_{c,s} \propto \langle \phi(r)\phi(-r) \rangle$ will be proportional to $K\cdot K^*$, thus the coefficient $b_{\nu}=Re[K_{\nu}]-Im[K_{\nu}]$. At the same time the local terms $a_{c,s} \propto \langle \phi(r)\phi(-r) \rangle$ will be proportional to $K^{-2}$, thus $a_{\nu}$ coefficients do acquire imaginary parts that do not cancel out.  

This has led to the new concept of a non-Hermitian TLL, defined by $H_{eff}$ with a complex-valued $K_{\rho}$ parameter. We are going to apply this result now. The only modification is that in our case the loss $\gamma(x)$ is space-dependent, namely it is non-zero only at the wires' end. When both real and imaginary parts of $K_{\rho}$ are varied, it is always possible to tune them in a way that ensures the condition $b_c\approx 2$ is fulfilled. It is our convention for $K_{\nu}$ parameters, as declared in Eq.\ref{eq:ham-TLL}, that makes it easier to spot this cancellation.

It is sensible to assume that both interaction and hybridization terms are $\propto N_{emit}$. To fulfil the condition $b_c=2$, we ought to assume that the TLL parameter $K_{\rho}$ not only changes close to the boundary, but also proportionally acquires an imaginary part. This can be justified by the fact that causality must be obeyed also by the emitted electrons, thus response functions such as susceptibility $\chi(\omega)$ must obey Kramers-Kroning relations (KKR), the relations that link integrals of real and imaginary components of $\chi(\omega)$. We are modifying the real part of charge susceptibility, $Re[\chi_{\rho}(\omega)]$, essentially by adding a part that peaks close to $q\approx 0$ -- interactions thus retain their long-range character. From KKR, this implies that necessarily the imaginary part $Im[\chi_{\rho}(\omega)]$ also changes. Precisely, there will be a double-peaked structure added in the imaginary part $Im[\chi_{\rho}(\omega)]$. This imaginary (i.e. finite loos) part corresponds to a tunneling, finite conductivity, into the plasmon-polaron state.  

\begin{figure}
    \centering
    \includegraphics[width=0.47\textwidth]{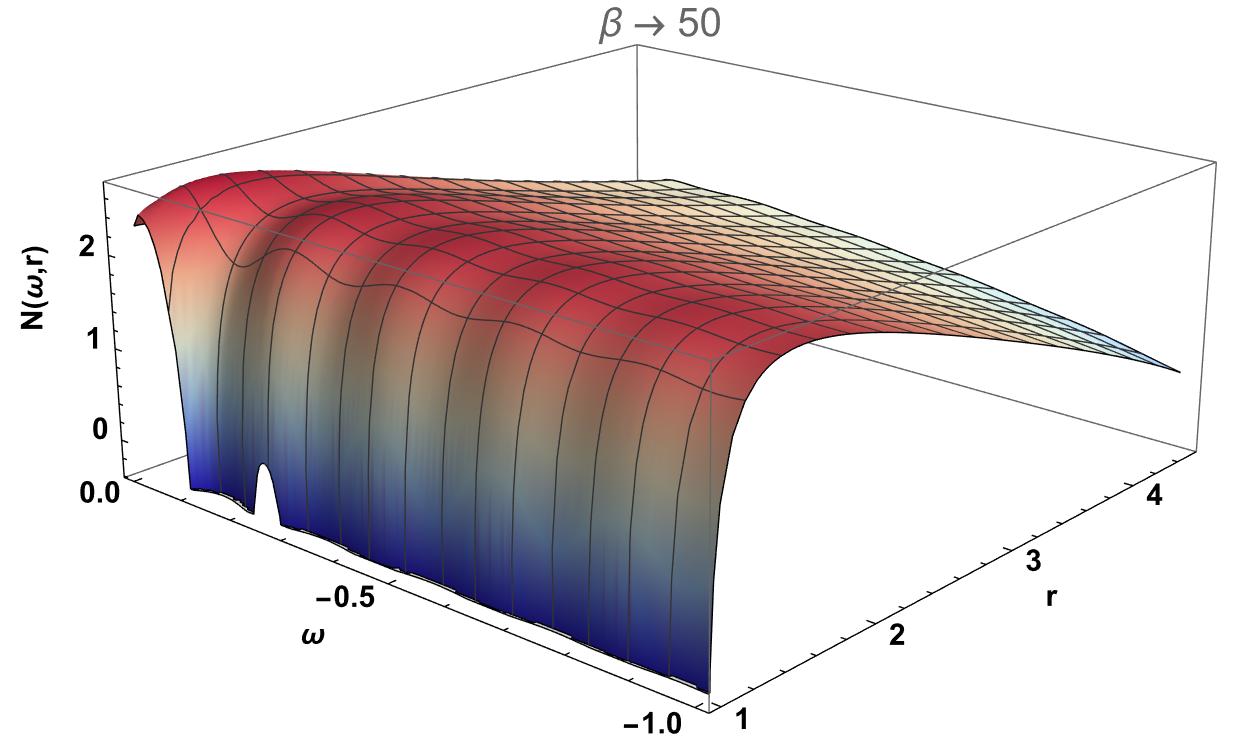}\\[0.2 cm]
    (a)\\[0.3 cm]
    \includegraphics[width=0.47\textwidth]{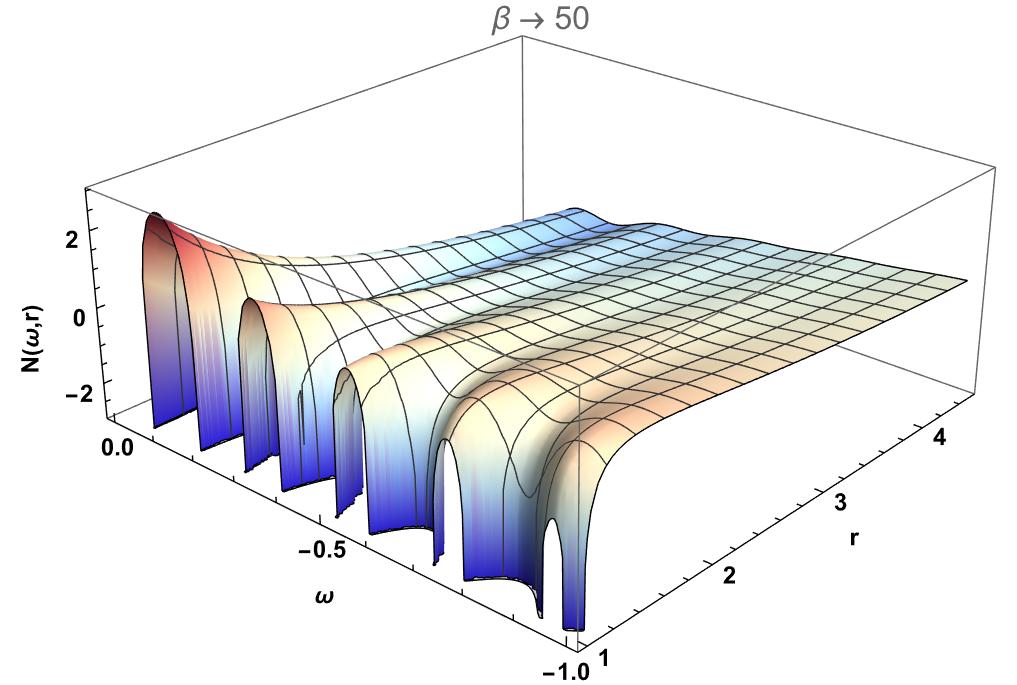}\\[0.2 cm]
    (b)
    \caption{LDOS for the case when $K_{\rho}$ varies in the vicinity of the boundary and also acquires a finite imaginary part therein. Top panel: Coulomb metal case (only $K_{\rho}\ll1$), and bottom panel: Hund metal case (both $K_{\rho}\ll1$ and $K_{\sigma}\ll1$).}
    \label{fig:Lindbladian}
\end{figure}

The plots of resulting LDOS with a substitution $K_{\rho}\rightarrow K_{\rho}(r) + \imath \gamma(r)$ implemented, are shown in Fig. \ref{fig:Lindbladian}. Both for the Coulomb and Hund case, we observe a small change of slope (due to the modified TLL parameter), and there is also a change of interference pattern (due to modified $v_{\rho}(r)$). For the Coulomb case, a weak interference pattern appears, while for the Hund case, the previously detectable beating between spin and charge modes is now blurred out.

 \begin{figure}
    \centering
    \includegraphics[width=0.45\textwidth]{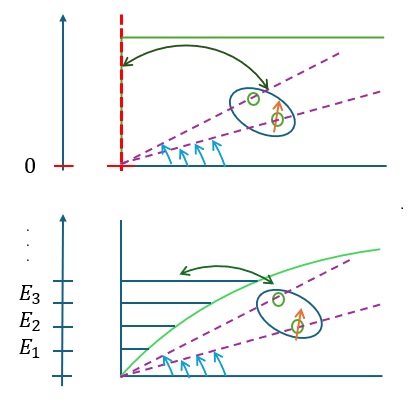}
    \caption{Schematic illustration of the boundary fermion hybridization with a TLL in Hund metal regime. Top: $K(r) = cste$, Bottom: $K(r) = 1/\epsilon(r)$. The top panel is the same as the bottom panel in Fig.\ref{fig:K-HundvsCoul}. The green line represents the Luttinger parameters, the dashed purple lines correspond to bosonic dispersions, and the blue arrows indicate thermal excitations. The arrow on the left side shows the energy spectrum on the boundary.}
    \label{fig: K(r)}
\end{figure}

The most remarkable difference is, however, for the Hund case where we previously observed a sharp resonant peak at zero energy. Now there are several broad peaks 
and the position of which has shifted towards a finite frequency. This phenomenon can be interpreted by the fact that now the radiation takes place into a 2D plasmon-polariton peak, which is a finite frequency feature. We have varied the $r^\varsigma$ dependence with changing exponent $\varsigma$ and we have checked different amplitudes of the real and imaginary parts (from $0$ to $0.5$), but the positions of the peaks essentially stay the same.

To make a connection with our previous argument: due to our boundary condition set on $\theta$ field, we are picking the odd solutions of the barrier problem. For Dirac-delta (i.e., strictly local $x=0$) boundary, there exists only $E_{0}=0$ solution in this class. When the 2D cloud of plasmon-polaritons allows for $K_{\rho}(r)$ dependence, which in the fermionic language is the finite size quantum well at the end of the 1D system, then we will have finite energy $E_0\neq 0$ solutions. This is illustrated in Fig.\ref{fig: K(r)}. In the top panel of Fig. \ref{fig: K(r)}, $K(r)=cste$, an infinitely sharp quantum well forms near the boundary, resulting in a hybridization. Here, a particle is created inside the TLL, where we combine a holon and a spinon to form a fermion-like object which can jump back and forth, resulting in an extra state.
In the bottom panel $K(r) = \frac{1}{\epsilon(r)}$, due to the spatial variation of $K(r)$ (we are changing the shape of the green line), there are discrete energy levels, producing resonances at finite energies, essentially the energy levels are appearing.   

From the point of view of Kondo-like problems, if we look at how the re-fermionization peak can be destroyed as we move away from the Emery-Kivelson point, as described in Ref.[\onlinecite{vonDelf-anisoKondo}], we see that in the presence of a finite energy scale coupled with boson density, called $\Delta_L$ therein, the peak indeed shifts to a finite frequency. In our case, this additional term in the Hamiltonian can be due to a finite distance over which the $K_{\rho}(r)$ varies and the boundary states induced therein, in the quantum well. In the sine-Gordon model, these boundary states are breathers (coupled states of solitons) whose energies are described as:

\begin{equation}\label{Eq: solitonic energy levels}
    E_{n}=\sqrt{\frac{2}{\pi K}}E_0\sin\left(\frac{n \pi K}{2(1-K)}\right)
\end{equation}
where $E_0$ comes from the solution of a boundary quantum well with $1/x^2$ potential. In Ref. [\onlinecite{Essin1x2}], it was shown that a self-adjoint solution with finite energy $E_0$ exists. 
For the Coulomb case, we have not invoked the Emery-Kivelson fixed point, so the above reasoning, involving the boundary fermion state, does not apply.

\begin{table}
\centering
\begin{tabular}{|c|c|c|}
\hline
  $E_i/E_j$ & From Eq. \ref{Eq: solitonic energy levels} & From Fig. \ref{fig:Lindbladian} (b) \\ 
   \hline
$E_2/E_1$ & 1.997 & 1.97 \\
$E_3/E_1$ & 2.990 & 3.02 \\
$E_4/E_1$ & 3.977 & 4.02 \\
$E_5/E_1$ & 4.954 & 5.11 \\
$E_3/E_2$ & 1.497 & 1.53 \\
$E_4/E_2$ & 1.991 & 2.04 \\
$E_5/E_2$ & 2.479 & 2.59 \\
$E_4/E_3$ & 1.329 & 1.33 \\
$E_5/E_3$ & 1.656 & 1.68 \\
$E_5/E_4$ & 1.245 & 1.26 \\
\hline
\end{tabular}
\caption{\label{tab: energy ratios} Comparison of energy level ratios received from Eq. \ref{Eq: solitonic energy levels} and observed values taken from Fig. \ref{fig:Lindbladian} (b).}
\end{table}

To test our boundary soliton-state conjecture one should compare the peak positions in Fig.\ref{fig:Lindbladian} with results of Eq.\ref{Eq: solitonic energy levels}. The only problem with using Eq. \ref{Eq: solitonic energy levels} directly is that it is quite hard to compute the exact value of $E_0$. However, it suffices to know that it is finite, to bypass this problem by considering the ratios of different energy levels and test our conjecture in this way. From the bottom panel of Fig. \ref{fig:Lindbladian}, we read out the energy values that correspond to the peaks and calculate their ratios. Comparing the results with the prediction of Eq.\ref{Eq: solitonic energy levels}, we find that the match is quite good as shown in table Tab.\ref{tab: energy ratios}; all the ratios indeed follow the equation for the solitonic masses (Eq. \ref{Eq: solitonic energy levels}). The only caveat is that for $E_1$ we have taken the value $E_1=0.18$, instead of $E_1^{obs}=0.12$ which we can read out from the figure. We suspect that this change $E_1^{obs}=2/3 E_1$ is because of the renormalization of the lowest energy level. This state is a single kink, a dipole as a solitonic state. When computing a local field acting on any dipole one needs to take into account a local depolarization field which is equal to $1/3$ of the bare polarization field \cite{griffiths2023introduction}. For higher energy states, we do not have dipoles, but we have quadrupoles, etc..., and the depolarization factors for these higher energy states are much smaller than for the dipole. Therefore, this extra renormalization is not present for $E_n$ when $n\geq 2$.

\section{Discussion and Conclusions}

\paragraph{Relation to other theories.} The main result of our work is that in the single-fermion spectral function (imaginary part of time-ordered Green's function) there is a zero-energy boundary state, but only in the case of Hund metal. In this case, both spin and charge degrees of freedom have a non-trivial spatial dependence close to the boundary. We interpret the peak as a manifestation of a Fresnel interference between two beating modes, a conjecture that is also supported by a corresponding modification of the temporal dependence of the scattering phase $\gamma(t)$. The existence of the boundary states has been also a subject of intense research within Cardy's formulation of  boundary CFT (in the string theory domain), see e.g. \text[\cite{CARDY1989581}]. It has been shown that such states, called Ishibashi states 
\cite{ishibashi1989boundary}, can exist, but only for selected cases when so-called gluing conditions can be fulfilled. These gluing conditions are expressed through derivatives 
\cite{rylands2020exact}, the Laurent mode currents. From this, we deduce that only when \emph{both} modes have spatial dependence (hence finite derivatives in both modes), the condition of a non-trivial cancellation between them at zero-energy can be fulfilled. In this context, it should be noted that our radiative (von Neumann) boundary condition is a difference between the two (left-/right-) Laurent currents, while a standard (Dirichlet) boundary condition corresponds to the cancellation of the sum of these currents. Both choices are equally valid from BCFT viewpoint but the peak will appear only in one of the two.  

The problem has been solved also directly as the boundary sine-Gordon model\cite{Zamolod-bound-sineG}. Also here authors confirmed existence of two types of boundaries either with or without surface resonance, the former present only in the interacting case when fixed field (not free field) boundary condition has been chosen (in the language of Sec.5 in Ref. [\onlinecite{Zamolod-bound-sineG}]). In our problem, we have several coupled copies of the boundary sine-Gordon problems. In this interpretation, the full fermion (that contains \emph{all} bosonic modes) has a boundary state only when boundary sine-Gordon models for \emph{all} modes fulfil the soliton-emitting-resonance condition. Overall, we ought to state that the existence of resonance is not a new finding of our study. What is remarkable is that we were able to recover it from interference between the TLL correlation functions.

\paragraph{Conclusions} The main outcome of this work is to obtain exact analytic results for LDOS profile for metallic, strongly correlated 1D systems at finite temperatures, the collective modes liquid known as Tomonaga-Luttinger liquid. Our results are important for a broad field of nanotechnology as the knowledge of the space-resolved DOS determines a distribution of valence electrons on the surface of any 1D or quasi-1D system. This is a directly measurable quantity, for instance, using STM or electron diffraction method. This quantity also determines the properties of nanostructures, for instance, a local conductivity of nano-circuits or a chemical reactivity on a surface of nanorods. In the case of such applications it is important to have the LDOS in a closed analytic formula, as it enables for its straightforward placement in further system modeling formulas.

In the paper we have obtained results for two types of TLLs: one is the Coulomb metal case, which has been already frequently recognized in the past, especially in the context of carbon-nanotubes type of materials (but also for p-orbital based sparse quasi-1D materials, like columnar chalcogenides). The other class is the Hund metal whose possible realization was identified as heavy atoms 1D wires, like $Au$ at stepped silicon surfaces. It should be noted that other realizations of Hund metal are also possible, especially based on the $K \leftrightarrow 1/K$ duality upon $\theta\leftrightarrow \phi$ change of the boundary condition. In this context, we notice that nanotubes based on heavy p-elements, such as rolled sheets of stanene, will have a strong local in-plane spin-orbit interaction. Then $K_{\sigma}\gg 1$, and the boundary condition is set by $\cos\phi_{\sigma}$. In this way, our framework, from Coulomb to Hund metals, spans all possible nanotubes based on p-elements hexagonal sheets.

\appendix

\section{Integration by parts}

The real space formula for LDOS, i.e., the imaginary part of the retarded single-particle fermionic propagator, as obtained by Eggert, has been expressed as:

\begin{equation}
    N(x,t)=\cos(\gamma(t))\bar{N}(x,t)
\end{equation}
where $\bar{A}(x,t)$ is the time-ordered correlation function of the single-particle fermionic propagator within TLL. It is the complicated function expressed in terms of $Sinh^a(x,t)$. The additional dependence that enters into the retarded correlation function, is introduced by a boundary phase shift, the $\cos(\gamma(t))$ where the angle $\gamma(t)$ is defined piece-wise: 
    
\begin{equation}
    \gamma(t) = 
\begin{cases} 
\frac{\pi}{2} (a_s + a_c), & 0 < t < \frac{2r}{v_c} \\
\frac{\pi}{2} (a_s + a_c + b_c), & \frac{2r}{v_c} < t < \frac{2r}{v_s} \\
\frac{\pi}{2} (a_s + a_c + b_c + b_s), & \frac{2r}{v_s} < t < \infty
\end{cases}
\end{equation}

and so the $\cos(\gamma(t))$ has a form of a sequence of step-functions as illustrated in Fig. \ref{fig: gamma}.

The overall time integral ($t$-domain Fourier transform to $\omega$-domain) can be solved by means of integration by parts, namely by taking $dv=\bar{N}(x,t)$ and $u=\cos(\gamma(t))\approx \sum \pm sign(t-t_i)$. Then we have: 
$$
\int_0^{\infty} u dv = - \int_0^{\infty} v du + v u\Bigg|_0^{\infty} 
$$
The first term simply gives $v(r,\omega,t)=\int \bar{N}(r,t;\omega)$, where the $t$-dependence is still there because the integral was indefinite. This is the integral that we present explicitly in Appendix B. Since $du$ is a sum of Dirac deltas (i.e., derivative of the sequence of step functions), then the first term reduces to $v(r,\omega,t=t_i)$ with $t_i(r)\propto r$. This is just a small correction, except in the case when $r\rightarrow 0$ where we reach singularity. The second term gives $v(r,\omega,t=\infty)\pm v(r,\omega,t=0)$ where the sign depends on the arrangement of $\cos(\gamma(t))$ which has a form of a sequence of step-functions as illustrated in Fig. \ref{fig: gamma}. We observe that the second term may cancel out the singularity of $v$ at zero (noticed in the first term); this happens for Coulomb metal. On the contrary for the Hund metal, there are two Dirac deltas of the opposite sign that cancel the first terms' singularity. We are then left with an unopposed constructive interference due to the second term that leads to an unsuppressed peak when $r\rightarrow 0$ and $\omega\rightarrow 0$. 

\begin{widetext}

\begin{figure}[t!]
    \centering
    \includegraphics[width=0.45\textwidth]{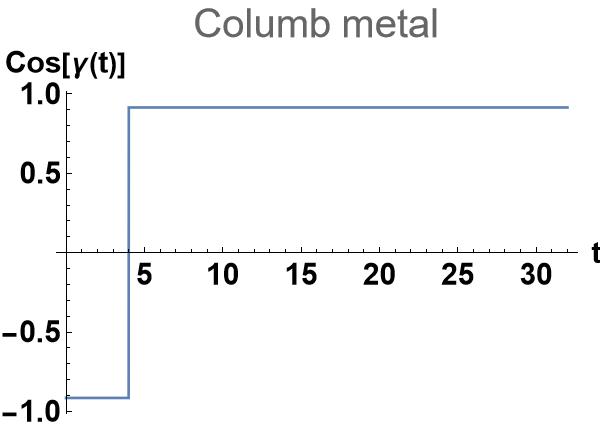} \hspace{1 cm}
    \includegraphics[width=0.45\textwidth]{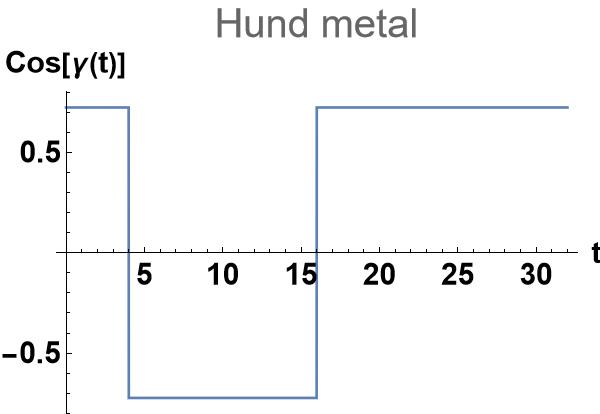}\\
     (a) \hspace{6.5 cm} (b)
    \caption{Time evolution of the boundary phase shift $\cos(\gamma(t))$ for (a) Columb metal and (b) Hund metal.  The function illustrates the induced modulations in the LDOS through boundary conditions within a TLL framework.}
    \label{fig: gamma}
\end{figure}


    
\section{Explicit form of a partial $\mathfrak{F}_{t\rightarrow\omega}$ Fourier transform}

Below we present results of integrals for the most general cases we have found. We obtained spectral function $N(\omega,\beta,r)$ at a finite frequency (measured like in STM as a distance from Fermi energy), finite temperature $T=\beta^{-1}$ and a finite distance $r$ from the end of a 1D system. We have bosonic modes propagating with two velocities $v_c$ and $v_s$ (note: dispersion may be degenerated and contain more than one mode) with corresponding exponents $a_{c,s}$ (see main text for their relation with the TLL parameters $K_{c,s}$). The result of the integral Eq. \ref{eq:boundTLL} for the case $-b_c/2 = 1$ and $-b_s/2 = 0$ (Coulomb metal) reads:

\begin{multline}
    N(\omega, \beta, r) =\frac{1}{2} \frac{v_c^{-a_c} v_s^{-a_s}}{ \pi^2 \alpha_{cut-off}}  \Bigg[ \pi^{a_c+a_s-1} \csch^2\left(\frac{2 \pi r}{v_c \beta}\right)  \beta \left( \frac{\beta }{\alpha_{cut-off}}\right)^{-a_c-a_s} e^{\frac{\pi t}{\beta}}  \sinh^{1-a_c-a_s}\left(\frac{\pi t}{\beta}\right) \\
    \times \Bigg[ 2 \cosh\left(\frac{4 \pi r}{\beta v_c}\right)
    \Biggl( \frac{ e^{-i \omega t} \, _2F_1 \left[ 1, \frac{1}{2} \left( 2-\frac{i \beta \omega}{\pi}-a_c-a_s \right),\frac{1}{2} \left( 2-\frac{i \beta \omega}{\pi}+a_c+a_s \right), e^{\frac{2 \pi}{\beta}t} \right] } {\frac{i \beta \omega}{\pi} - a_c -a_s  } \\
    -\frac{ e^{i \omega t}  \, _2F_1 \left[ 1, \frac{1}{2} \left( 2+\frac{i \beta \omega}{\pi}-a_c-a_s\right), \frac{1}{2} \left( 2+\frac{i \beta \omega}{\pi}+a_c+a_s \right), e^{\frac{2 \pi }{\beta}t} \right] } {\frac{i \beta \omega}{\pi} + a_c + a_s  } \Biggl)\\
    +  \frac{ e^{-i \omega t} e^{-\frac{2 \pi t}{\beta}} \, _2F_1 \left[ 1, -\frac{1}{2} \left(\frac{i \beta  \omega}{\pi} + a_c + a_s \right), \frac{1}{2} \left(-\frac{ i \beta  \omega}{\pi} + a_c + a_s \right), e^{\frac{2 \pi }{\beta}t}\right] } {-2 -\frac{i \beta \omega }{\pi}+ a_c + a_s  }  \\
    + \frac{ e^{-i \omega t} e^{\frac{2 \pi t}{\beta}} \, _2F_1 \left[ 1, \frac{1}{2} \left( 4-\frac{i \beta \omega}{\pi}-a_c-a_s\right), \frac{1}{2} \left( 4-\frac{i \beta \omega}{\pi}+a_c+a_s\right), e^{\frac{2 \pi }{\beta}t}\right] } {2 - \frac{i \beta \omega}{\pi} + a_c + a_s} \\
    +\frac{e^{i \omega t} e^{-\frac{2 \pi t}{\beta}} \, _2F_1  \left[  1, \frac{1}{2} \left(-\frac{i \beta  \omega}{\pi} + a_c + a_s \right), \frac{1}{2} \left(\frac{i \beta  \omega}{\pi} + a_c + a_s \right), e^{\frac{2 \pi}{\beta}t} \right]  } {-2 + \frac{i \beta \omega}{\pi} + a_c + a_s  }  \\
    + \frac{ e^{i \omega t} e^{\frac{2 \pi t}{\beta}} \, _2F_1 \left[ 1, \frac{1}{2} \left( 4+\frac{i \beta \omega}{\pi}-a_c-a_s\right), \frac{1}{2} \left( 4+\frac{i \beta \omega}{\pi}+a_c+a_s\right) , e^{\frac{2 \pi t}{\beta}}\right]   } {2 +\frac{i \beta \omega}{\pi} + a_c + a_s  } \Bigg]\\
    - t \left(\frac{t}{\alpha_{cut-off}}\right)^{-a_c-a_s} \Biggl( \frac{t^2 v_c^2}{r^2 (a_c+a_s-3)}-\frac{4}{a_c+a_s-1} \Biggl) \Bigg]\Bigg|_{t=t_{ins}}^{t=t_{life}}
\end{multline}\label{eq:Coulumb}

The result of integral Eq. \ref{eq:boundTLL} for the case $-b_c/2 = 1$ and $-b_s/2 = 1$ (Hund metal) reads:

\begin{multline}
    N(\omega, \beta, r) = \frac{1}{8} \frac{ v_c^{-a_c} v_s^{-a_s} }{\pi^2 \alpha_{cut-off} }\Biggl[  \pi^{a_c+a_s-1} \csch^2\left(\frac{2 \pi r}{v_c \beta}\right)  \csch^2\left(\frac{2 \pi r}{v_s \beta}\right)
    \beta  \left( \frac{\beta }{\alpha_{cut-off} }\right)^{-a_c-a_s} \\ \times e^{\frac{\pi t}{\beta}}  \sinh^{1-a_c-a_s}\left(\frac{\pi t}{\beta}\right) \Biggl[ 4 \left( \frac{1}{2} - \cosh\left(\frac{4 \pi r}{\beta v_c }\right) \cosh\left(\frac{4 \pi r}{\beta v_s }\right) \right) \\
    \times \Biggl( \frac{e^{-i \omega t} \, _2F_1 \left[ 1, \frac{1}{2} \left( 2-\frac{i \beta \omega}{\pi}-a_c-a_s \right),\frac{1}{2} \left( 2-\frac{i \beta \omega}{\pi}+a_c+a_s \right), e^{\frac{2 \pi t}{\beta}} \right] } {\frac{i \beta \omega}{\pi} - a_c - a_s  } \\
    -\frac{ e^{i \omega t}  \, _2F_1 \left[ 1, \frac{1}{2} \left( 2+\frac{i \beta \omega}{\pi}-a_c-a_s\right), \frac{1}{2} \left( 2+\frac{i \beta \omega}{\pi}+a_c+a_s \right), e^{\frac{2 \pi t}{\beta}} \right] } {\frac{i \beta \omega}{\pi} + a_c + a_s  }\Biggl) \\
    -  \Bigg(\frac{ e^{-i \omega t} e^{\frac{-4 \pi t}{\beta}} \, _2F_1 \left[ 1, -\frac{1}{2} \left(2+\frac{i \beta  \omega}{\pi} + a_c + a_s \right), \frac{1}{2} \left(-2-\frac{ i \beta  \omega}{\pi} + a_c + a_s \right), e^{\frac{2 \pi t}{\beta}}\right]}{-4 - \frac{i \beta \omega}{\pi} + a_c + a_s  }  \\
    + \frac{ e^{-i \omega t} e^{\frac{4 \pi t}{\beta}}\, _2F_1 \left[ 1, \frac{1}{2} \left( 6-\frac{i \beta \omega}{\pi}-a_c-a_s\right), \frac{1}{2} \left( 6-\frac{i \beta \omega}{\pi}+a_c+a_s\right), e^{\frac{2 \pi t}{\beta}}\right] } {4 - \frac{i \beta \omega}{\pi} + a_c + a_s} \\
    + \frac{e^{i \omega t} e^{\frac{-4 \pi t}{\beta}} \, _2F_1  \left[  1, \frac{1}{2} \left(-2+\frac{i \beta  \omega}{\pi} - a_c - a_s \right), \frac{1}{2} \left(-2+\frac{i \beta  \omega}{\pi} + a_c + a_s \right), e^{\frac{2 \pi t}{\beta}} \right]  } {-4 + \frac{i \beta \omega}{\pi} + a_c + a_s  }  \\
    + \frac{e^{i \omega t} e^{\frac{4 \pi t}{\beta}}\, _2F_1 \left[ 1, \frac{1}{2} \left( 6+\frac{i \beta \omega}{\pi}-a_c-a_s\right), \frac{1}{2} \left( 6+\frac{i \beta \omega}{\pi}+a_c+a_s\right) , e^{\frac{2 \pi t}{\beta}}\right]   } {4 +\frac{i \beta \omega}{\pi} + a_c + a_s  } \Bigg) \\
    -  \left( e^{-\frac{4 \pi r}{\beta v_c}} + e^{\frac{4 \pi r}{\beta v_c}} +e^{-\frac{4 \pi r}{\beta v_s}} +e^{-\frac{4 \pi r}{\beta v_s}} \right)  \\
    \times  \Biggl( \frac{  e^{-i \omega t } e^{\frac{2 \pi t}{\beta}} 
   \, _2F_1 \left[ a_c+a_s, \frac{1}{2} \left( 2 - \frac{i \beta \omega}{\pi}+a_c+a_s\right), \frac{1}{2} \left( 4-\frac{i \beta \omega}{\pi}+a_c+a_s\right) , e^{\frac{2 \pi t}{\beta}}\right]   } {2 - \frac{i \beta \omega}{\pi} +  a_c +  a_s } \\
   + \frac{e^{-i \omega t } e^{\frac{-2 \pi t}{\beta}}  \, _2F_1  \left[ a_c+a_s , \frac{1}{2} \left(-2-\frac{i \beta  \omega}{\pi} + a_c + a_s \right), \frac{1}{2} \left(-\frac{i \beta  \omega}{\pi} + a_c + a_s \right), e^{\frac{2 \pi t}{\beta}} \right]  } {-2 -\frac{i \beta \omega }{\pi} + a_c + a_s  }\\
    + \frac{ e^{i \omega t } e^{-\frac{2 \pi t}{\beta}} \, _2F_1  \left[ a_c+a_s , \frac{1}{2} \left(-2+\frac{i \beta  \omega}{\pi} + a_c + a_s \right), \frac{1}{2} \left(\frac{i \beta  \omega}{\pi} + a_c + a_s \right), e^{\frac{2 \pi t}{\beta}} \right]  } {-2 + \frac{i \beta \omega}{\pi} + a_c + a_s  }\\
    + \frac{ e^{i \omega t } e^{\frac{2 \pi t}{\beta}} \, _2F_1  \left[ a_c+a_s , \frac{1}{2} \left(2+\frac{i \beta  \omega}{\pi} + a_c + a_s \right), \frac{1}{2} \left(4+\frac{i \beta  \omega}{\pi} + a_c + a_s \right), e^{\frac{2 \pi t}{\beta}} \right]  } {2 + \frac{i \beta \omega }{\pi}+ a_c + a_s  } \Biggl) \Bigg] \\
    - t \left(\frac{t}{\alpha }\right)^{-a_c-a_s} \left(-\frac{t^4 v_c^2 v_s^2}{r^4 \left(a_c+a_s-5\right)}+\frac{4 t^2 \left(v_c^2+v_s^2\right)}{r^2 \left(a_c+a_s-3\right)}-\frac{16}{a_c+a_s-1}\right)\Biggl]\Bigg|_{t=t_{ins}}^{t=t_{life}}
\end{multline}\label{eq:Hunds}

In the above formula, we explicitly kept the dependence on the time limit of the integral. In the main text, we identified it as the incompleteness of the Beta function which corresponds to setting finite limits of the time window over which one integrates. This is the same situation as in Ref. [\onlinecite{chudzinski2020contribution}]. The finite temporal limits have their physical interpretation: i) the upper limit represents the finite lifetime of bosonic particles $t_{life}$ that is present in an open system; ii) the lower limit represents either energy cut-off of the theory $\sim 1/\alpha_{cut-off}$ of the finite time-span of a single instanton $t_{ins}$ event at the 1D system's end (depending which one is longer characteristic time). Keeping them in the expression allows us to incorporate them into the modelling of a given physical realization of an open TLL.


\end{widetext}

\section{Testing special cases of $\Delta K_{\rho}(r)$}

Since in Fig. \ref{fig:Lindbladian} we take the real and imaginary parts as equal, then the condition of $b_c=2$ remains fulfilled, but if we take only the imaginary part of $K_\rho$ then our solution might not apply.  Taking purely real or imaginary $\Delta K$ is an artificial construct with no connection to TLL. The aim of these calculations is basically to test our hypothesis about the origin of the peaks.

\begin{figure}[H]
    \centering
    \includegraphics[width=1\linewidth]{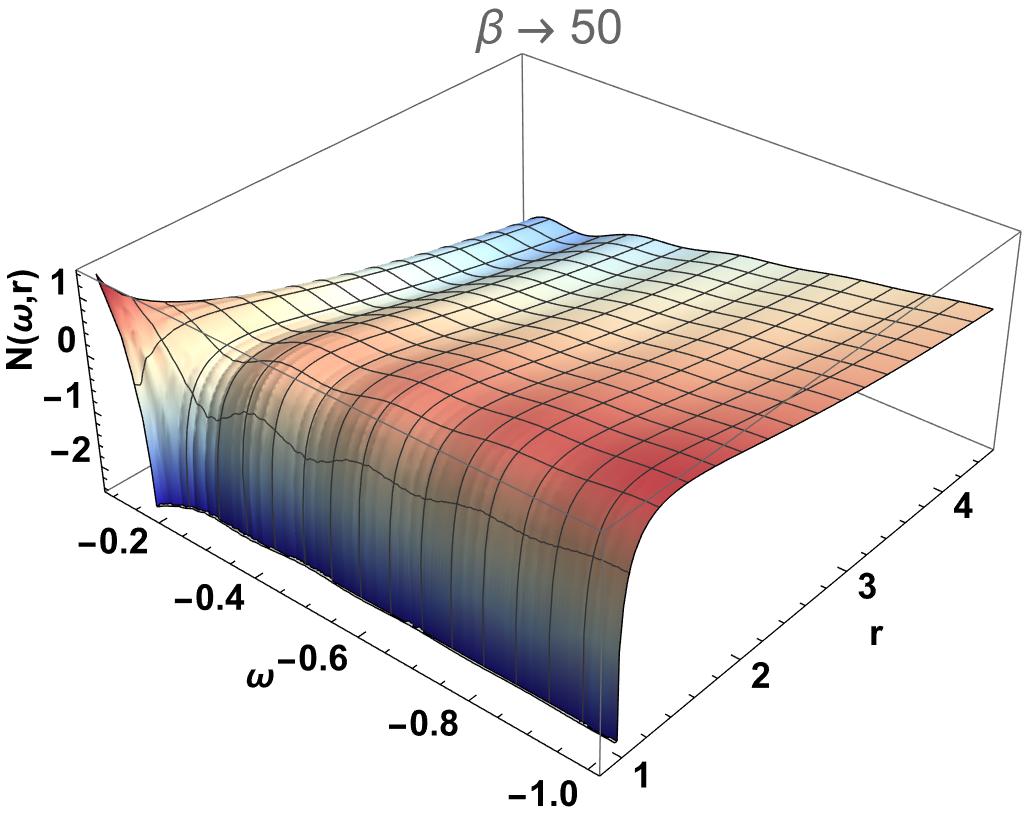} \\[0.2 cm]
    \caption{LDOS for Hund case, as in the bottom panel of Fig. \ref{fig:Lindbladian} but containing only the imaginary part of modified $K_{\rho}(r)$, the $i\gamma(r)$.}
    \label{fig: Hund_imaginary}
\end{figure}

\paragraph{Pure imaginary $\gamma(r)$} Fig. \ref{fig: Hund_imaginary} illustrates the LDOS for Hund metal, where $K_\rho(r)$ is with only the imaginary part $\gamma(r)$ is included. By removing the real part, through the comparison with Fig.\ref{fig:Lindbladian}b) (where both real and imaginary parts are included), we can extract the role played by the real part of $\Delta K_{\rho}(r)$, the part that is now missing. One immediately recognizes that the peaks close to $r\rightarrow 0$ are now absent. Therefore, the conjecture made in the bottom panel of Fig. \ref{fig: K(r)},  proves that the real part of $K(r)$ dependence plays a role in the emergence of these states.

\begin{figure}[H]
    \centering
    \includegraphics[width=1\linewidth]{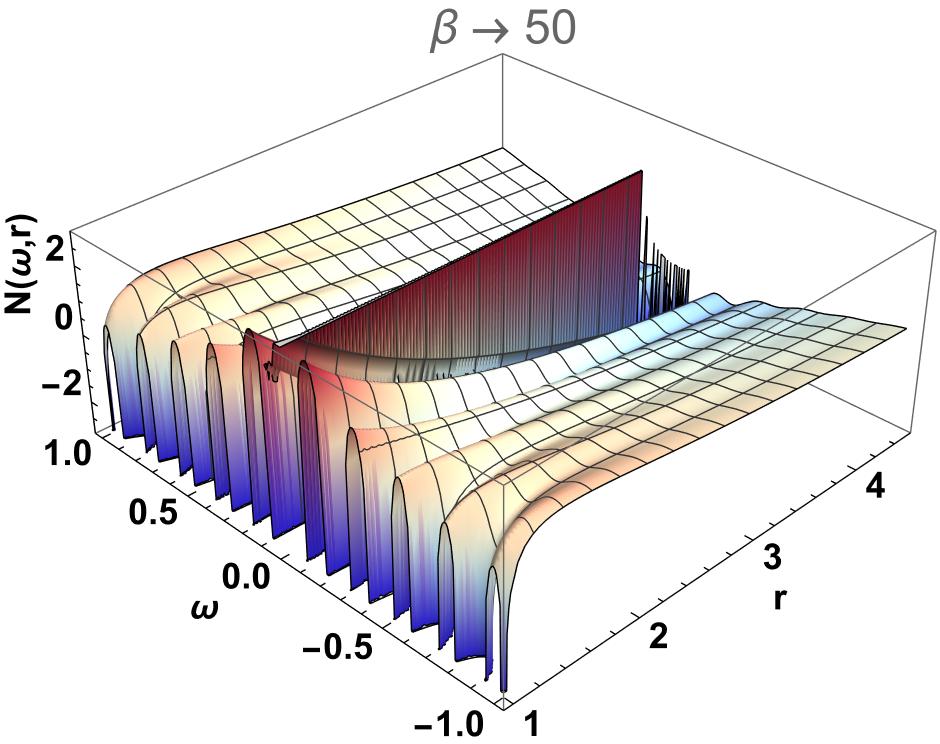}\\[0.2 cm]
    \caption{LDOS for Hund case, as in the bottom panel of Fig. \ref{fig:Lindbladian} but containing only the real part of $\Delta K_{\rho}(r)$.}
    \label{fig: Hund_real}
\end{figure}

\paragraph{Purely real $\Delta K$.} In an analogous way we now take the case without imaginary $\gamma(r)$. The result is presented in Fig.\ref{fig: Hund_real}. Now we observe the finite frequency resonances (as caused by the $Re[K_{\rho}(r)]$ dependence) but also the zero frequency peak. The zero frequency peak has been interpreted as being due to the oddness of BC. In the presence of both real and imaginary parts, the situation considered in the main text, we have BC condition with a twist where parity symmetry is broken \cite{zawadzki2017symmetries}. Then oddness is not well defined and the central peak is missing. Once we remove the imaginary part we restore the symmetry and the zero frequency peak appears again.

\bibliography{biblio.bib}

\end{document}